\documentclass[aps,prd,preprint,nofootinbib]{revtex4-1}
\pdfoutput=1

\usepackage[utf8]{inputenc} 
\usepackage{amsmath}
\usepackage{amssymb}
\usepackage{xcolor}
\usepackage{graphicx}
\usepackage{dcolumn}
\usepackage{bm}
\usepackage{bbm}
\usepackage{mathtools}
\usepackage[colorlinks=true]{hyperref}
\usepackage{physics}
\usepackage{comment}
\usepackage[normalem]{ulem}
\usepackage{braket}

\newcommand{\Hphi}{\mathcal{H}_\varphi}
\newcommand{\Hp}{\mathcal{H}_P}
\newcommand{\Op}{\mathcal{O}}
\newcommand{\Pphi}{ {\bf P}_\varphi }
\newcommand{\p}{ {\bf p} }
\newcommand{\kb}{ {\bf k} }
\newcommand{\q}{ {\bf q} }
\newcommand{\x}{ {\bf x} }
\newcommand{\y}{ {\bf y} }
\newcommand{\Lb}{ {\bf L} }
\newcommand{\vg}{ {\bf v} }

\begin{document}

\title{Quantum Dynamics of Probe Particles in Thermal Fields \\ and the Emergence of Stochastic Dynamics}

\author{Bruno Scheihing-Hitschfeld}
\email{bscheihi@kitp.ucsb.edu}
\affiliation{Kavli Institute for Theoretical Physics, University of California, Santa Barbara,
California 93106, USA}

\preprint{INT-PUB-26-033}

\begin{abstract}
    Starting from a general Hamiltonian describing the quantum dynamics of probe particles interacting with a set of environment degrees of freedom, we derive the quantum master equation with which the reduced density matrix of a probe particle evolves, the general equilibration condition that they satisfy when the environment is prepared in a thermal state --- encoded in a KMS property for line operators extended in real time --- and the Langevin description that emerges.
    The novelty of our results resides in their generality: We do not assume a specific form of the environment self-coupling, of the environment operator that couples the particle to it, of the statistics of its correlation functions, whether the statistics of the momentum transfer is Gaussian or not, or whether the probe particles move relativistically. Our results only require that spacetime translations, parity and time reversal be symmetries of the Hamiltonian, and that there exists a separation of scales between those characterizing the dispersion relation of the probe particle and those of the environment. We also discuss the limiting case in which Gaussian Brownian motion emerges. Overall, our results constitute a first-principles derivation of what properties of the underlying quantum field theory govern the energy loss and momentum fluctuations of probe particles.
\end{abstract}

\maketitle

\newpage

\tableofcontents

\newpage

\section{Introduction}

Understanding how dynamical macroscopic behavior emerges from the underlying dynamics of a large number of microscopic degrees of freedom is one of the main goals of contemporary physics. A paradigmatic example of the emergence of macroscopic behavior is the Brownian motion of a point(-like) particle immersed in a large thermal environment. At sufficiently long times, the coarse-grained dynamics of such a particle is described by a Langevin equation
\begin{equation}
    \frac{d p_i }{dt} = \xi_i \, , \label{eq:Langevin}
\end{equation}
where $p_i$ is the momentum of the particle, and $\xi_i$ is a stochastic force that emerges as a coarse-grained (i.e., long-time) result of many microscopic interactions with the other degrees of freedom in the system.\footnote{It is common practice to this equation as $\frac{d p_i }{dt} = F_i + \xi_i$, where $F_i$ is the mean force the particle experiences (a common example is a drag force $F_i = -\eta_D p_i$), and $\xi_i$ describes the fluctuations around this mean. We have kept all contributions inside $\xi_i$ because our discussion will treat them on equal footing.}

The theoretical study of the Brownian motion that this equation describes, and of the stochastic force that characterizes it, goes back to the works of Einstein, Sutherland, von Smoluchowski and Langevin~\cite{Sutherland1905,Einstein1905,vSmoluchowski1906,Langevin1908}. The stochastic force $\xi$, and consequently the ``probe'' particle's motion, encodes properties of the interaction of the probe particle with the medium and of the medium itself. If an explicit connection between the microscopic physics and the stochastic force can be established, this dynamics can be used to gain a deeper understanding of the underlying microscopic theory of the medium.

Many derivations of Eq.~\eqref{eq:Langevin} exist, ranging from the foundational works~\cite{Sutherland1905,Einstein1905,vSmoluchowski1906,Langevin1908} to much more general versions in the context of classical statistical mechanics with extensions to quantum dynamics over the past century~\cite{Zwanzig1973NonlinearGL,Mori,NordholmZwanzig,zwanzig2001nonequilibrium,BreuerPetruccione}. A paradigmatic example is the Caldeira-Leggett model~\cite{Zwanzig1973NonlinearGL,CALDEIRA1983374,CaldeiraPhysRevB.48.13974}, where the emergence of a Langevin equation comes directly from a particle-bath Lagrangian where the bath is described by a collection of harmonic oscillators.\footnote{See also~\cite{Petrosyan:2021lqi} for a relativistic version of this model.}

Quantum Brownian motion is a problem of intrinsic interest in theoretical physics. It makes appearances in a broad class of physical systems, from electrons in solids to heavy quarks in quark-gluon plasma. The latter is a particularly interesting example, as it features physics of fluctuations and dissipation of a heavy particle interacting with an environment where \textit{all} of the degrees of freedom at play are fundamental fields of the Standard Model of Particle Physics. As such, it simultaneously constitutes an example of complex emergent behavior and a direct test of our understanding of the dynamics of the fundamental degrees of freedom of matter. This and many other situations make it invaluable to have theoretical techniques that accommodate fields as degrees of freedom. In addition to core quantum mechanics techniques, formal methods to describe this problem in field theory include the Schwinger-Keldysh path integral~\cite{Schwinger:1960qe,Keldysh:1964ud} and the Feynman-Vernon influence functional~\cite{FEYNMAN1963118}.\footnote{See also~\cite{Erdos2010arXiv1009.0843E} for a discussion based on the Schr\"odinger equation.}

That being said, derivations of the Langevin equation starting from the microscopic Lagrangian density governing a given physical system, where the stochastic force is explicitly characterized by quantities defined and (in principle) calculable in the microscopic theory, have largely relied on simplifying assumptions (e.g., assuming the resulting stochastic force $\xi$ is Gaussian).  
While there have been efforts using the Schwinger-Keldysh path integral to characterize the non-Gaussian features that can appear in the resulting stochastic description in specific models~\cite{HuPazZhangPhysRevD.45.2843,HuPazZhangPhysRevD.47.1576,Lin:2023bli,Cho:2026prp,Chakrabarty:2018dov,Chakrabarty:2019qcp}, it is desirable to establish model-independent properties that hold for non-Gaussian, and possibly relativistic stochastic systems where the fluctuations have a quantum origin. In particular, without making assumptions regarding whether the dynamics of the environment is weakly or strongly coupled.

A class of models that has provided significant guidance to understand Brownian motion in strongly coupled quantum systems is that of theories that possess a holographic dual via the AdS/CFT correspondence~\cite{Maldacena:1997re,Witten:1998qj}. Starting from the realization that heavy quarks in quark-gluon plasma could be modeled as Brownian particles~\cite{vanHees:2004gq,Moore:2004tg,vanHees:2005wb}, shortly thereafter it was proposed that holography provides a meaningful model of heavy quarks propagating in such a strongly coupled plasma~\cite{Herzog:2006gh,Gubser:2006bz,Casalderrey-Solana:2006fio,Gubser:2006nz,
Casalderrey-Solana:2007ahi}, where the transport coefficients are directly calculable in the microscopic theory of the thermal bath. From this quantum mechanical starting point, a stochastic picture of the dual string was subsequently pursued~\cite{deBoer:2008gu,Son:2009vu,Giecold:2009cg,Casalderrey-Solana:2009ifi,Caron-Huot:2011vtx} and later investigations looked more closely into the non-Gaussian~\cite{Chakrabarty:2019aeu} and the relativistic~\cite{Bu:2021jlp} features of this process. However, in all of that time, Gubser's observation~\cite{Gubser:2006nz} that the Einstein relation was badly broken at relativistic speeds\footnote{In this theory, the longitudinal momentum diffusion coefficient scales as $\gamma^{5/2}$ and the drag force as $\gamma$, where $\gamma = (1-v^2/c^2)^{-1/2}$ is the Lorentz boost factor of the heavy quark in the rest frame of the medium.} remained puzzling. Only much more recently it was pointed out via an explicit calculation that a simultaneous treatment of non-Gaussian and relativistic features of the stochastic force are crucial in order to have self-consistent heavy quark equilibration dynamics to a Boltzmann distribution~\cite{Rajagopal:2025ukd}. 
Further work revealed that the analytic property that guarantees equilibration in this setup is, in fact, independent of the details of the gauge theory to which the heavy quark is coupled~\cite{Rajagopal:2025rxr}. It stands to reason to ask: does this property hold more generally?

In this work, we derive the Langevin equation~\eqref{eq:Langevin} in a systematic, theory-agnostic way, starting from first principles in quantum mechanics and statistical mechanics, generalizing recently found insights for quantum gauge field theories~\cite{Rajagopal:2025rxr}, without making use of Gaussian, weak/strong coupling or non-relativistic assumptions. The main result of this work is to provide an explicit link between a certain correlation function of line operators (to be introduced in the course of our discussion) in the microscopic quantum theory of a given system, generated by its interaction with the aforementioned probe particle, and the emergent stochastic force that characterizes the dynamics of the probe particle. This result makes explicit:
\begin{enumerate}
    \item The equilibration condition that the process satisfies, together with the KMS-type condition that guarantees it.
    \item The generic appearance of non-Gaussian statistics in the stochastic force, and in which limit it can be well approximated by a Gaussian process.
    \item How the microscopic, reversible quantum dynamics gives rise to irreversible, dissipative stochastic (classical) motion as a consequence of the statistical average over the environment degrees of freedom.
\end{enumerate}
As in the case where the coupling to the environment is weak, Markovian evolution is a consequence of the large separation of scales between the correlation time of the environment and the time it takes to substantially alter the probe particle's momentum. In our case, we shall focus on probe particles with dispersion relations characterized by an energy/time scale that is well separated from the scales of the medium. A concrete example is a particle with a mass much larger than the temperature of its environment. Because the separation of time scales is provided in this way, there is no assumption needed regarding whether the environment is weakly or strongly coupled.\footnote{While our discussion will make ample use of this scale separation later on in this work, we highlight that many systems do not possess such separation and non-Markovian dynamics appears as the generic case~\cite{BreuerPetruccione}. Scale-invariant quantum systems are generically non-Markovian~\cite{Arguelles:2026bah}.
Even the same physical system can cease to undergo Markovian dynamics depending on the circumstances. For example, while the dynamics of single heavy quarks as introduced above can be thought of as Markovian, heavy quark-antiquark bound states in quark-gluon plasma need not be Markovian~\cite{Yao:2021lus,Nijs:2023dks,Nijs:2023dbc,BR:2025lhx,Kim:2026cyu} because the binding energy levels can be of the same order as the energy scales of the environment.}

Our results provide a direct link between the fundamental considerations that lead to the Fluctuation-Dissipation theorem~\cite{Callen1951,Kubo1966} and the physics of far-from-equilibrium probe particles. In fact, they encode a new fluctuation-dissipation relation between the zero-frequency limit of elements of a family of $n$-point correlation functions of local operators. While there is a rich and substantial literature on generalized fluctuation-dissipation relations~\cite{Bernard1959,Peterson1967,Haag1967,efremov1969fluctuation,stratonovich1970contribution,Bochkov1981,Chou1985,hao1981closed,Carrington:1996rx,Wang:1998wg,Miyazaki2005,Dubkov2009} that go beyond the original version of the theorem, a derivation of how non-Gaussian stochastic motion of a probe particle emerges from first principles, without making simplifying assumptions regarding the nature of the surrounding medium, had not been given before.

This paper is organized as follows: In Section~\ref{sec:setup} we describe the conceptual ingredients of our setup and introduce the general form of the Hamiltonian that we will study throughout. In Section~\ref{sec:quantum} we give our operational definition of what ``probe particle'' means in terms of a condition at the level of the aforementioned Hamiltonian, derive general expressions for the quantum transition amplitudes that quantify the probabilities for the probe particle to go between different momentum states, and derive a quantum master equation for the evolution of the reduced density matrix describing the momentum states of the particle. These constitute the specific setup and provide the conceptual and notational framework in which our main result becomes manifest.
The main result of this work is a general derivation of the KMS condition for line operators, which we present in Eq.~\eqref{eq:W-KMS}, and the implication that a kinetic equilibrium solution with Boltzmann weights~\eqref{eq:equil-solution} follows from it. This is presented in Section~\ref{sec:KMS}. It can be understood as a field-theoretic derivation of the detailed balance condition a probe particle coupled to a thermal environment satisfies. Section~\ref{sec:stochastic} then discusses the consequences of this result: how a (classical) stochastic process emerges and how its time evolution operator is related to fundamental quantities in the underlying field theory, the tower of fluctuation-dissipation relations between $n$-point functions of the environment correlators that follows from Eq.~\eqref{eq:W-KMS}, and how a Gaussian stochastic process may emerge. We discuss applications and future directions in Section~\ref{sec:outlook}. Appendix~\ref{app:large-t-derivation} supplements the discussion of the derivation of the master equation.

\section{Setup} \label{sec:setup}

We consider a quantum theory endowed with a Hilbert space composed of two subsystems,
\begin{equation}
    \mathcal{H} = \Hp \otimes \Hphi \, , \label{eq:Hilbert-fact}
\end{equation}
where $\Hp$ is the Hilbert space of the probe particle in isolation, and $\Hphi$ the Hilbert space that describes the medium through which it is moving.

The space $\Hp$ is characterized by a vacuum state $\ket{0}$ and creation/annihilation operators $a^\dagger_i(\p)$, $a_i(\p)$ satisfying $a_i(\p) \ket{0} = 0$ and commutation relations
\begin{align}
    [a_i(\p),a^\dagger_j(\q)]_{\pm} = (2\pi)^d \delta_{ij} \delta^{(d)}(\p - \q) \, , & & [a_i(\p) , a_j(\q)]_{\pm} = 0 \, ,
\end{align}
where $[\cdot,\cdot]_{\pm}$ denotes a commutator if the subscript is a plus sign, and an anti-commutator if it is a minus sign, corresponding to the Hilbert space of a free boson and a free fermion, respectively. We display both because the statistics of the probe particle will not be crucial in what follows. What is crucial is that we assume the existence of an unbounded continuous label $\p \in \mathbb{R}^d$ that we identify with the momentum of the particle in each state. The extra label $i$ denotes any additional discrete degrees of freedom that the particle might have, such as spin or color\footnote{There are additional considerations that need attention in theories where the color symmetry is gauged and interactions proceed via the gauge field, including the fact that the physical Hilbert space ceases to be factorizable as in~\eqref{eq:Hilbert-fact}. We will not go into such details here; for discussion on some of these points (in the context of heavy quark effective theory, where this is crucial) see~\cite{Rajagopal:2025rxr,Lin:2026nlr}.}. We also define
\begin{align}
    a_i(\x) \equiv \int_{\p} e^{i \p \cdot \x } a_i(\p) \, ,
\end{align}
where we have introduced the shorthand $\int_{\p} \equiv \int \tfrac{d^dp}{(2\pi)^d}$, which corresponds to an annihilation operator localized at a \textit{spatial} position $\x$. For integrals over $\x$ we will use the shorthand $\int_{\x} \equiv \int d^d x$.

On the other hand, the only requirement we will impose on the space $\Hphi$ is that there exists an energy operator (Hamiltonian) $H_\varphi$ and a conserved momentum operator $\Pphi$ (i.e., satisfying $[H_\varphi,\Pphi] = 0$) that generates spatial translations
\begin{equation}
    e^{-i \Pphi \cdot \y } \phi(\x) e^{i \Pphi \cdot \y } =  \phi(\x+\y) \label{eq:translations}
\end{equation}
for local operators $\phi(\x)$ (i.e., supported at the point in space specified by its argument $\x$) acting on $\Hphi$. We shall assume the existence of these operators, but we will not need to specify an explicit form for them.

To fully define the physical system, we need to specify the Hamiltonian for the complete system. We introduce it as
\begin{equation}
    H = \int_\p E(\p) \, a_i^\dagger(\p) a_i(\p) \otimes \mathbbm{1}_\varphi + \int_{\x}  a_i^\dagger(\x) a_j(\x) \otimes \Op_{ij}(\x) + \mathbbm{1}_P \otimes H_\varphi \, , \label{eq:Hamiltonian}
\end{equation}
where $\Op_{ij}$ is an operator on the environment degrees of freedom that obeys Eq.~\eqref{eq:translations} in the place of $\phi$.\footnote{Note that $\Op_{ij}$ need not be a local operator from the point of view of the theory describing the environment degrees of freedom. For example, if $\phi(\x)$ is a local operator, then for any function $f(\x)$,
\begin{equation}
    \Op(\x) = \int_\y f(\x - \y) \phi(\y) \nonumber
\end{equation}
satisfies Eq.~\eqref{eq:translations} and is nonlocal.
}
This Hamiltonian satisfies three properties that will be crucial later on:
\begin{enumerate}
    \item It conserves the total probe particle number:
    \begin{align}
        [H, N_P] = 0 \, , \quad \quad {\rm with} \quad \quad N_P = \sum_i \int_\p a_i^\dagger(\p) a_i(\p) \, ,
    \end{align}
    \item the non-interacting energy of the probe particles depends only on their momentum $\p$, but not on the internal labels $i$ (that is to say, these labels enjoy a symmetry in the free theory), and
    \item all the operators in the interaction term have the same position argument, i.e., it is a local interaction from the point of view of the probe.
\end{enumerate}
Finally, although this will only become relevant much later, we will also assume that parity and time reversal are symmetries of this theory.

Equation~\eqref{eq:Hamiltonian} is the starting point of all our subsequent discussion. In short, it describes otherwise free particles with a prescribed dispersion relation $E(\p)$ interacting locally with an environment, where all interactions that may change their number are deemed to be negligible.\footnote{It should be noted that, while one can study the dynamics of a relativistic particle in this way by setting $E(\p) = \sqrt{\p^2 + M^2}$, the Hamiltonian~\eqref{eq:Hamiltonian} does not describe a Lorentz-invariant theory for the probe particles. Such a theory would generically also include interactions with the environment that create two particles $\propto \Op a^\dagger a^\dagger$ or annihilate two particles $\propto \Op a a$ (mathematically, this is because local operators in relativistic theories contain both creation and annihilation operators; also, when written in terms of $a,a^\dagger$ they also contain other kinematic factors that make the Hamiltonian different from~\eqref{eq:Hamiltonian}). However,
\vadjust{\vskip\maxdimen}
if the mass $M$ of said particles is large, such processes become suppressed in the limit of low energy/momentum transfer, and the leading terms in a $1/M$ expansion of $H$ reproduce exactly the form that we use later~\eqref{eq:Hamiltonian-approx} (in fact, this is the heavy quark effective theory expansion~\cite{Georgi:1990um}).} Multiparticle interactions that conserve number, of the schematic form $a^\dagger a^\dagger a a$ (or with a higher number of $a$'s) would play no role in our later analysis as we shall restrict ourselves to the sector of the Hilbert space with only one probe particle.

\section{Quantum dynamics of a probe particle}
\label{sec:quantum}

To study the dynamics of the probe particle in the environment described by $\Hphi$, we consider an initial state given by a density matrix
\begin{equation}
    \rho = \rho_P \otimes e^{-\beta H_\varphi} \, ,
\end{equation}
where we take $\rho_P$ to be supported in a neighborhood around some central value $\p$ of the momentum that characterizes the state. Concretely,
\begin{equation}
    \rho_P = \int_{\kb,\q} \rho^\p_{ij}(\kb,\q) \ket{\p+\kb,i} \bra{\p+\q,j} \, , \label{eq:init-generic}
\end{equation}
where $\rho^\p_{ij}(\kb,\q)$ is a function localized around $({\bf 0},{\bf 0})$. The indices $i,j$ are the internal labels of the probe particle states. Loosely speaking, we require $|\kb|,|\q| \ll |\p|$. The precise requirement for our derivation to hold will become clear later.

In simpler words, we will study the quantum dynamics of a particle with a characteristic momentum given by $\p$. 

As in any quantum theory, we may study the dynamics in terms of probability amplitudes. In fact, for our purposes the elementary ``amplitude'' is
\begin{equation}
    \mathcal{A}_{(\kb_i,j) \to (\kb_f,i)}^\p \equiv \braket{ \p + \kb_f, i | U(t) | \p + \kb_i, j }  \, , \label{eq:amplitude-def}
\end{equation}
where we have used quotation marks because this amplitude is actually an operator on $\Hphi$. Here, $U(t) = e^{-iHt}$ is the time evolution operator of the full theory, with both the probe and environment included.

For example, a quantity of interest calculable from this definition~\eqref{eq:amplitude-def} and the initial state~\eqref{eq:init-generic} is the probability for the probe particle to lose momentum $\kb$ after propagating for a time $t$. For concreteness, setting $\rho_P = \ket{\p} \bra{\p}$, and omitting the internal label indices, this probability is
\begin{align}
    P(\kb;\vg,t) &= \frac{1}{Z} {\rm Tr}_{\mathcal{H} } \! \left[ \mathcal{A}_{0 \to -\kb}^\p e^{-\beta H_\varphi} \left(\mathcal{A}_{0 \to -\kb}^\p \right)^* \right] \nonumber \\
    &= \frac{1}{Z} {\rm Tr}_{\mathcal{H}} \! \left[ \ket{ \p - \kb } \braket{ \p - \kb | U(t) | \p } \bra{\p} \otimes e^{-\beta H_\varphi}  U^\dagger(t)  \right] \, , \label{eq:prob-k-loss}
\end{align}
the meaning of which can be straightforwardly interpreted as time-evolving the initial density matrix and projecting onto the probe particle state with momentum $\p - \kb$. The normalization $Z$ is simply the trace of the initial state, $Z = {\rm Tr}_{\mathcal{H}} \left[ \rho_P \otimes e^{-\beta H_\varphi} \right]$.

In order to characterize the dynamics, the task we need to carry out is to use the Hamiltonian~\eqref{eq:Hamiltonian} to evaluate quantities like Eq.~\eqref{eq:prob-k-loss}.
To make progress, we may profit from the fact that we are interested in specific states. Concretely, we may expand the ``free'' part of the Hamiltonian~\eqref{eq:Hamiltonian} around the characteristic momentum $\p$ of our states of interest, obtaining 
\begin{equation}
    H = \int_\kb \left[E(\p) + \vg \cdot \kb + \ldots \right] \, a_i^\dagger(\p+\kb) a_i(\p+\kb) \otimes \mathbbm{1}_\varphi + \int_{\x}  a_i^\dagger(\x) a_j(\x) \otimes \Op_{ij}(\x) + \mathbbm{1}_P \otimes H_\varphi \, , \label{eq:Hamiltonian-expansion}
\end{equation}
where we have introduced the probe particle (group) velocity $\vg \equiv \partial E/\partial \p$, and ``$\ldots$'' stands for terms that are higher order in $\kb$, which are small as long as 
\begin{equation}
    E(\p+\kb) \approx E(\p) + \vg \cdot \kb \, . \label{eq:kinematic-approx}
\end{equation}
That is to say, if the interaction with the environment does not alter the momentum of the probe to an extent that Eq.~\eqref{eq:kinematic-approx} becomes invalid, we may use 
\begin{equation}
    \tilde{H} = \int_\kb \left[E(\p) + \vg \cdot \kb \right] \, a_i^\dagger(\p+\kb) a_i(\p+\kb) \otimes \mathbbm{1}_\varphi + \int_{\x}  a_i^\dagger(\x) a_j(\x) \otimes \Op_{ij}(\x) + \mathbbm{1}_P \otimes H_\varphi \, , \label{eq:Hamiltonian-approx}
\end{equation}
as the generator of time evolution. We take the validity of this approximation as the criterion that defines whether the particle is a ``probe'' of the system: if this approximation were not valid, it means that the interactions are strong enough to appreciably change the momentum of the particle on time scales comparable to those of the environment dynamics, in which case the particle will rarely remember enough of its initial condition to serve as a probe.

We emphasize that Eq.~\eqref{eq:Hamiltonian-approx} can only describe the dynamics starting from probe particle states with momentum close enough to $\p$ --- and that this is consistent with Eq.~\eqref{eq:init-generic}. Whether this approximation eventually breaks down is thus a dynamical question --- this breakdown can only happen if time evolution makes it so that the momentum change in the probe particle is large --- and a calculation has to be carried out in order to assess this quantitatively. In principle, one would like to compare the exact result of the full theory~\eqref{eq:Hamiltonian} with the approximate one~\eqref{eq:Hamiltonian-approx}. In practice, however, a self-consistency check in the approximate theory is often the best one can hope for.\footnote{We will return to this point later in subsection~\ref{sec:extending}.}

The reason why it is convenient to use the Hamiltonian~\eqref{eq:Hamiltonian-approx} becomes apparent when one writes all of the creation/annihilation operators in terms of the position coordinate, allowing one to collect together the free and interacting terms of the probe particle as
\begin{equation}
    \tilde{H} = \int_\x (e^{i \p \cdot \x} a_i^\dagger(\x) ) \left[(E(\p) - i \vg \cdot \nabla_\x) \delta_{ij} \otimes \mathbbm{1}_\varphi  + \mathbbm{1}_P \otimes \Op_{ij}(\x) \right]  (e^{-i \p \cdot \x} a_j(\x)) + \mathbbm{1}_P \otimes H_\varphi \, . \label{eq:Hamiltonian-approx-position}
\end{equation}

This is a substantial simplification because the equations of motion for the creation and annihilation operators $a_i(\x,t)$ and $a^\dagger_i(\x,t)$ are explicitly solvable in terms of the operator $\mathcal{O}_{ij}$ that couples them to the thermal bath. In what follows, the time argument of all operators reflects time evolution generated by $\tilde{H}$, i.e., for an operator $A$ we have $A(t) = U^\dagger(t) A(0) U(t)$ with $U(t) \equiv e^{-iH_\varphi t}$ and $A(0) \equiv A$. Explicitly, the Heisenberg picture equation of motion
\begin{align}
    \frac{\partial a_i(\x,t)}{\partial t} &= i [\tilde{H} , a_i(\x,t)] \nonumber \\
    &= - i e^{i\p \cdot \x} \left[(E(\p) - i \vg \cdot \nabla_\x) \delta_{ij} \otimes \mathbbm{1}_\varphi  + \mathbbm{1}_P \otimes \Op_{ij}(\x,t) \right]  (e^{-i \p \cdot \x} a_j(\x,t)) \, ,
\end{align}
is solved by
\begin{equation}
    a_i(\x,t) = e^{-i(t-t_0)[E(\p) - \p \cdot \vg]}  \left[ {\rm P} \exp \left( -i \int_{t_0}^t dt' \mathcal{O}(\x - (t-t') \vg  ,t') \right) \right]_{ij} a_j(\x - (t-t_0)\vg,t_0) \, , \label{eq:operator-sol}
\end{equation}
where ${\rm P}$ stands for a path-ordering symbol (equivalent to time-ordering in this case) prescribing how operators are ordered in the exponential, exactly as in Dyson's formula~\cite{Dyson:1949bp}. This applies both to operator ordering and to the ``matrix multiplication'' over the internal indices $i,j$. For notational simplicity, we denote
\begin{equation}
    W^{\vg,\x}_{ij}(t,t_0) \equiv e^{-i(t-t_0)[E(\p) - \p \cdot \vg]}  \left[ {\rm P} \exp \left( -i \int_{t_0}^t dt' \mathcal{O}(\x + t' \vg  ,t') \right) \right]_{ij} \, . \label{eq:W-op-def}
\end{equation}

It follows that the transition amplitude between two momentum eigenstates of the (free) probe particle after a time $t$ has passed is
\begin{align}
    \mathcal{A}_{(\kb_i,j) \to (\kb_f,i)}^\p &= \int_{\y,\x} e^{-i( \p + \kb_f )\cdot \y} e^{i( \p + \kb_i )\cdot \x} \braket{0 | a_i(\y,0) U(t) a_j^\dagger(\x,0) | 0 } \nonumber \\
    &= \int_{\y,\x} e^{-i( \p + \kb_f )\cdot \y} e^{i( \p + \kb_i )\cdot \x} \braket{0 | U(t) a_i(\y,t) a_j^\dagger(\x,0) | 0 } \nonumber \\
    &= \int_{\y,\x} e^{-i( \p + \kb_f )\cdot \y} e^{i( \p + \kb_i )\cdot \x} e^{-i H_\varphi t} \braket{0 | W^{\vg,\y-\vg t}_{ij}(t,0) a_i(\y-\vg t,0) a_j^\dagger(\x,0) | 0 } \nonumber \\
    &= \int_{\y,\x} e^{-i( \p + \kb_f )\cdot \y} e^{i( \p + \kb_i )\cdot \x} e^{-i H_\varphi t} \braket{0 | W^{\vg,\y - \vg t}_{ij}(t,0) | 0 } \delta^{(d)}(\x-\y+\vg t) \nonumber \\
    &= e^{-i( \p + \kb_f )\cdot \vg t} e^{-i H_\varphi t} \int_\x e^{-i (\kb_f - \kb_i) \cdot \x } \braket{0 | W^{\vg,\x}_{ij}(t,0) | 0 } \, ,
\end{align}
which, as advertised earlier, is an operator on $\Hphi$.

Furthermore, because the Hamiltonian $\tilde{H}$ conserves probe particle number, the projection onto the probe particle vacuum $\braket{0 | W^{\vg,\x}_{ij}(t,0) | 0 }$ enforces that the operator $\Op_{ij}$ in $W^{\vg,\x}_{ij}(t,0)$ may be evolved using only the environment Hamiltonian $H_\varphi$ instead of the full $\tilde{H}$. (Every time $\tilde{H}$ acts on $\ket{0}$, only $H_\varphi$ remains. Since $a_i(\x,0)$ commutes with $H_\varphi$, acting with $\tilde{H}$ multiple times on $\ket{0}$ will still yield an equivalent outcome to acting on it with $H_\varphi$ the same number of times.) In fact, it is straightforward to show that
\begin{equation}
    \braket{0 | W^{\vg,\x}_{ij}(t,0) | 0 } = e^{-i[E(\p) - \p \cdot \vg]t} e^{i H_\varphi t} e^{-i \Pphi \cdot \vg t } \left[\exp \left( - i H_\varphi t + i \Pphi \cdot \vg t - i \Op(\x,0 ) t \right) \right]_{ij} \, , \label{eq:W-op-def-expli}
\end{equation}
making it explicit that the degrees of freedom of the probe do not appear in the amplitude $\mathcal{A}_{(\kb_i,j) \to (\kb_f,i)}^\p$.

In summary, we have obtained an explicit expression for the transition amplitudes of the probe particle
\begin{equation}
    \mathcal{A}_{(\kb_i,j) \to (\kb_f,i)}^\p = e^{-i[ E(\p) + \kb_f \cdot \vg ] t} e^{-i \Pphi \cdot \vg t } \int_\x e^{-i (\kb_f - \kb_i) \cdot \x }  \big[\exp \left( - i [H_\varphi - \Pphi \cdot \vg + \Op(\x,0 )] t \right) \big]_{ij} \, ,
\end{equation}
fully in terms of the environment degrees of freedom. As before, the indices of this operator should be understood as the indices of the exponentiated operator $\Op$.
These provide the essential ingredients for the evolution equations we shall derive in what follows.

\subsection{A closer look at the environment}

Before proceeding to derive our main results, it is instructive to examine the properties of the operator that defines the amplitude $\mathcal{A}_{(\kb_i,j) \to (\kb_f,i)}^\p$. Namely, the operator that encodes the interaction of the probe and the environment is
\begin{equation}
    \tilde{U}(t) \equiv \exp \left( - i [H_\varphi - \Pphi \cdot \vg + \Op(\x,0 )] t \right) \, , \label{eq:total-unitary}
\end{equation}
which we have denoted as $\tilde{U}(t)$ due to its explicit unitary nature. If it were not for the operator $\Op$ that mediates the interaction with the probe, the eigenstates of this operator would be exactly those of $H_\varphi$. With the interaction taken into account, the interpretation of the eigenstates of this operator is rather natural: they are the eigenstates of the ``boosted'' medium (with ``free'' time evolution operator $H_\varphi - \Pphi \cdot \vg$) in the presence of a point test particle situated at position $\x$.
In theories with Lorentzian invariance under boosts this connection can be made even more apparent, as this invariance implies that the spectrum of the operator $H_\varphi - \Pphi \cdot \vg$ is the same as that of $\gamma^{-1} H_\varphi$, with $\gamma = (1 - |\vg|^2)^{-1/2}$ the Lorentz boost factor.\footnote{We note that the combination $H_\varphi - \Pphi \cdot \vg$ appearing in $\tilde{U}$ can be written covariantly $\int d\Sigma_\nu \beta_\mu T^{\mu \nu}$. The same is true for the instance of $H_\varphi$ appearing in the thermal density matrix. We discuss some aspects of writing these operators in different reference systems in Section~\ref{sec:relativistic}.}

The problem of studying the amplitudes $\mathcal{A}_{(\kb_i,j) \to (\kb_f,i)}^\p$ then gets exactly mapped to the problem of understanding the spectrum of $\tilde{U}(t)$, i.e., the eigenstates of the environment in the presence of the test particle.

\subsection{Master equation for the reduced density matrix of the probe particle}

We are now in a position to return to the dynamics of the probe particle in the presence of a thermal environment. Starting from the initial condition~\eqref{eq:init-generic}, by keeping track of the probe particle degrees of freedom and tracing over the environment, we obtain that the reduced density matrix after a time $t$ is
\begin{align}
    \rho^\p_{il}(\kb_f,\q_f;t) &\equiv e^{i(\kb_f-\q_f) \cdot \vg t} \braket{\kb_f,i | {\rm Tr}_{\Hphi}[\rho(t)] | \q_f, l} \nonumber \\
    &= e^{i(\kb_f-\q_f) \cdot \vg t} \frac{1}{Z} \int_{\kb_i,\q_i} \rho^\p_{jk}(\kb_i,\q_i) \, {\rm Tr}_{\Hphi} \! \left[ \mathcal{A}_{(\kb_i,j) \to (\kb_f,i)}^\p e^{- \beta H_\varphi} \left(\mathcal{A}_{(\q_i,k) \to (\q_f,l)}^\p\right)^\dagger \right] \nonumber \\
    &= \frac{1}{Z} \int_{\kb_i,\q_i,\x,\y} e^{-i(\kb_f-\kb_i)\cdot \x} e^{i(\q_f-\q_i)\cdot \y}   \rho^\p_{jk}(\kb_i,\q_i) \, {\rm Tr}_{\Hphi} \! \left[ W^{\vg,\x}_{ij}(t,0) e^{- \beta H_\varphi} W^{\vg,\y}_{kl}(0,t) \right] \, , \label{eq:rho-evol-full}
\end{align}
where we have used that $(W^{\vg,\x}_{ij}(t,t_0))^\dagger = W^{\vg,\x}_{ji}(t_0,t)$, and we have included a phase $e^{i(\kb_f-\q_f) \cdot \vg t}$ in the definition of $\rho^\p_{il}$ for future convenience. Furthermore, from here on we have omitted the projection of $W^{\vg,\x}_{ij}$ onto the state with zero probe particles, and as such the $\Op$ operators in all instances of $W$ must be understood as being evolved only with the environment Hamiltonian $H_\varphi$.

It is clear that all of the nontrivial information about the dynamics is encoded in
\begin{equation}
    \langle W^{\vg,t}_{il,jk} \rangle (\y-\x) \equiv \frac{1}{Z} {\rm Tr}_{\Hphi} \! \left[ W^{\vg,\x}_{ij}(t,0) e^{- \beta H_\varphi} W^{\vg,\y}_{kl}(0,t) \right] \, . \label{eq:eff-evol-def}
\end{equation}
We shall refer to this object as the ``evolution loop.''
Because $[H_\varphi,\Pphi] = 0$, this expression in fact only depends on the difference between the spatial coordinates $\x$ and $\y$. Consequently, we may carry out two of the integrals in Eq.~\eqref{eq:rho-evol-full} to obtain
\begin{equation}
    \rho^\p_{il}(\kb_f,\q_f;t) = \int_{\kb_i, \Lb} e^{i ( \kb_f - \kb_i ) \cdot \Lb}  \rho^\p_{jk}(\kb_i, \kb_i + (\q_f - \kb_f) ) \langle W^{\vg,t}_{il,jk} \rangle (\Lb) \, . \label{eq:evloution-result-inbin}
\end{equation}

Incidentally, but not coincidentally, if one averages over the internal label of the probe particle, the Fourier transform of the evolution loop is exactly the (averaged) momentum change probability $P(\kb;\vg,t)$ that we introduced earlier in Eq.~\eqref{eq:prob-k-loss}:
\begin{equation}
    P(\kb;\vg,t) = \int_{\Lb} e^{-i \kb \cdot \Lb } \langle \sum_{i,j} W^{\vg,t}_{ii,jj} \rangle (\Lb) \, , \label{eq:P-of-W-averaged}
\end{equation}
making the meaning of $\langle W^{\vg,t}_{il,jk} \rangle (\Lb)$ a bit more transparent.

So far, all of the steps we have carried out have followed as mathematical consequences of the starting point. To proceed further, we need to draw input from physical considerations on the dynamics of the environment degrees of freedom governed by $H_\varphi$, so as to establish the properties that $\langle W^{\vg,t}_{il,jk} \rangle (\Lb)$ may satisfy. Indeed, the generic situation for the long-time limit (relative to the environment time scales) of this expectation value is that a \textit{Markovian} regime emerges, where the evolution loop satisfies a local-in-time evolution equation
\begin{equation}
    \partial_t \left[ \langle W^{\vg,t}_{il,jk} \rangle (\Lb) \right] = -\left[ S_{il,mn}(\Lb,\vg) + O(t^{-1}) \right] \langle W^{\vg,t}_{mn,jk} \rangle (\Lb) \, . \label{eq:markovian}
\end{equation}
See Appendix~\ref{app:large-t-derivation} for a discussion of how this property emerges from a path integral point of view (in particular, how Markovianity emerges as a consequence of scale separation).

It is then straightforward to derive that $\rho^\p_{il}$ also satisfies a Markovian differential equation in time
\begin{align}
    \partial_t \rho^\p_{il}(\kb_f,\q_f;t) &= - \int_{\kb_i, \Lb} S_{il,mn}(\Lb,\vg) e^{i ( \kb_f - \kb_i ) \cdot \Lb}  \rho^\p_{jk}(\kb_i, \kb_i + (\q_f - \kb_f) ) \langle W^{\vg,t}_{mn,jk} \rangle (\Lb) \nonumber \\
    &= -  \int_{\kb, \Lb} S_{il,mn}(\Lb,\vg)  e^{-i \kb \cdot \Lb}  \rho^\p_{mn}(\kb_f + \kb, \q_f + \kb; t ) \, , \label{eq:m-derivation}
\end{align}
fully characterized by $S_{il,mn}(\Lb,\vg)$. This is the master equation that governs the evolution of the probe particle density matrix.

A striking feature of this equation is that it only couples entries of the density matrix that have the same momentum difference between its two entries. That is to say, the diagonal momentum entries $\rho^\p_{il}(\kb,\kb;t)$ are only coupled to other entries of the same form, and more generally, entries of the form $\rho^\p_{il}(\kb,\kb + \Delta \kb;t)$ are only coupled to other entries with the same value of $\Delta \kb$. In fact, introducing a redefined version of the reduced density matrix,
\begin{equation}
    \tilde{\rho}^\p_{il}({\bf k}, \Delta {\bf k};t) \equiv \rho^\p_{il}\left( \kb + \frac12 \Delta \kb,\kb - \frac12 \Delta \kb;t \right) \, ,
\end{equation}
their evolution equations may be immediately written as
\begin{equation}
    \partial_t \tilde{\rho}^\p_{il}({\bf k}, \Delta {\bf k};t) = - S_{il,mn}(-i \partial_{\kb} ,\vg) \tilde{\rho}^\p_{mn}({\bf k}, \Delta {\bf k};t) \, , \label{eq:master-final-single-v}
\end{equation}
with the same form of the dynamics for every value of $\Delta \kb$, all of which are mutually decoupled. It is pleasing that $\tilde{\rho}^\p_{il}$ appears naturally in the form that enters the Wigner transform~\cite{Wigner:1932eb}, foretelling that some of the equations we will later obtain describe the dynamics of the phase space distribution of the probe particle. It is also pleasing to see that a series expansion of $S$ in powers of its first argument (which can also be carried out in all of the subsequent versions of the master and kinetic equations we shall present later on) yields the Kramers-Moyal expansion~\cite{Kramers1940,Moyal1949}.

\subsection{Extending the master equation to arbitrary probe particle momentum} \label{sec:extending}

The previous derivation relied on assuming that the density matrix is nonzero only for probe particle momenta ${\bf p} + {\bf k}$ sufficiently close to ${\bf p}$, so that they would all approximately correspond to the same velocity ${\bf v}$ and the dynamics is well-described by the approximated version of the Hamiltonian $\tilde{H}$. However, at sufficiently late times, it is completely conceivable (and in fact generic) that the support of $\tilde{\rho}^\p_{ij}$ evolved with Eq.~\eqref{eq:master-final-single-v} will eventually move outside the domain of applicability of the approximation $E({\bf p}+{\bf k}) \approx E({\bf p}) + {\bf v} \cdot {\bf k}$.

That being said, there is a natural way to substantially ameliorate this problem. Going through the steps that led to Eq.~\eqref{eq:m-derivation}, there is no obstruction whatsoever to promoting the arguments of the density matrix to be \textit{total} momenta (instead of a small deviation around ${\bf p}$) provided that their difference is still small so that both may be described by one velocity. The crucial step is to note that the ${\bf k}_i$ integration variable in Eq.~\eqref{eq:m-derivation} should also modify the velocity of the particle ${\bf v} = \vg(\p) \to \vg(\p+\kb_i) $ when the domain of the integral is all momenta. Then, in terms of the reduced density matrix of the probe particle \textit{without} assuming it is fully localized around $\p$, but rather only around the diagonal 
\begin{equation}
    \rho_P(t) = \int_{\p, \Delta \kb} e^{-i \Delta \kb \cdot \vg(\p) t } \tilde{\rho}_{ij}(\p, \Delta \kb;t) \ket{\p +  \Delta \kb/2,i} \bra{\p -  \Delta \kb/2 ,j } \, 
\end{equation}
[where, consistent with localization around the diagonal of $\rho_P$, $\Delta \kb$ must be such that $E({\bf p}+ \Delta{\bf k}/2) \approx E({\bf p}) + {\bf v} \cdot \Delta {\bf k}/2$], then Eq.~\eqref{eq:m-derivation} becomes
\begin{align}
    \partial_t \tilde{\rho}_{il}(\p,\Delta \kb;t) &= -  \int_{\kb, \Lb} S_{il,mn}(\Lb,\vg(\p + \kb )  )  e^{-i \kb \cdot \Lb}  \tilde{\rho}_{mn}(\p + \kb, \Delta \kb; t ) \, , \label{eq:m-derivation-i}
\end{align}
which also gives rise to Markovian dynamics
\begin{equation}
    \partial_t \tilde{\rho}_{il}({\bf p}, \Delta {\bf k};t) = - S_{il,mn}(-i \partial_{\p} ,\vg(\p)) \tilde{\rho}_{mn}({\bf p}, \Delta {\bf k};t) \, , \label{eq:master-final}
\end{equation}
with one noteworthy difference from Eq.~\eqref{eq:master-final-single-v}: the partial derivative with respect to momentum in the first argument of $S_{il,mn}$ also acts on its second argument, lending itself explicit sensitivity to what the dispersion relation of the probe particle is (the visual effect that $\vg$ is left to the right of $\partial_\p$ is why we choose to keep $\vg$ as the second argument of $S$ instead of as a label).

Equation~\eqref{eq:master-final} is our final result at the level of a master equation for the reduced density matrix of the probe particle. It describes the leading-order dynamics of probe particles in the approximation where the momentum transfer they receive from the medium is small enough that their dispersion relation may be linearized at each individual time step, but at the same time that the time spent in the medium in each individual time step is sufficiently long for the dynamics to be Markovian.

To be more precise, assuming the momentum transfer probability spreads out from being supported at ${\bf k} = 0$ without large ``jumps'' (i.e., with a time-dependent bound for the probability outside a region containing the origin that goes to zero as $t \to 0$), there will always be a sufficiently short time for which the error generated by the linearization of the dispersion relation around $\p$ will be below any specified tolerance. However, depending on the properties of the medium, this time may not be long enough for the Markovian regime described by Eq.~\eqref{eq:markovian} to set in. In that case, before this late-time regime may be reached, there will be non-negligible contributions from higher-order terms in the expansion of the dispersion relation around $\p$. In practice, those will manifest as higher-order spatial derivatives than in Eq.~\eqref{eq:Hamiltonian-approx-position}, making the time evolution of the annihilation operators in Eq.~\eqref{eq:operator-sol} no longer described in terms of operators integrated along a line (hereafter ``line operators''). For example, if quadratic order terms in $\nabla_\x$ are included, then the solution in Eq.~\eqref{eq:operator-sol} would be modified to include a diffusive-like spread in position space, leading to a more complicated structure than just line integrals. It would, of course, be very interesting to characterize the dynamics including such corrections as well, but that lies outside the scope of our present discussion.

Nonetheless, leaving those caveats aside, we stress that the dynamics defined by Eq.~\eqref{eq:master-final}, which is characterized by the long-time limit of a correlation function of line operators, is self-consistent: while one should check that the simplifying approximations from the dynamics in the full quantum system are under control in order to, e.g., characterize a concrete property of $H_\varphi$ via measurements of the probe, the dynamics is well-defined and, at least in principle, does not run into any obstruction at late times. We will elaborate on this in the next section by examining the properties of the evolution operator $S$. Most importantly, we will see that this equation is guaranteed to have the thermal state among its steady states.

\section{The KMS relation for line operators}
\label{sec:KMS}

Having established the dynamics of the reduced density matrix of the probe particle, we now turn to examining the properties of the operator that defines it. 

According to Eq.~\eqref{eq:markovian}, in the late-time limit the evolution loop may be written as
\begin{equation}
    \langle W^{\vg,t}_{il,jk} \rangle (\Lb) = \exp \left( - t S(\Lb,\vg) + O(t^{0}) \right)_{il,jk} \, , \label{eq:W-exp-S}
\end{equation}
which specifies how the properties of $ \langle W^{\vg,t}_{il,jk} \rangle (\Lb)$ map to those of $S_{ij,jk}(\Lb,\vg)$ and vice-versa.

There is, in fact, a structural property that the evolution loop obeys, following only from its definition~\eqref{eq:eff-evol-def} and the definition of the line operator~\eqref{eq:W-op-def}, which generalizes the derivation presented in~\cite{Rajagopal:2025rxr} for the case of heavy quarks in gauge theories. For notational convenience in the next few steps, we introduce
\begin{equation}
    U_{ij}[(t_f,{\bf x}_f),(t_i,{\bf x}_i)] \equiv \left[ {\rm P} \exp \left( -i \int_{t_i}^{t_f} dt' \mathcal{O} \! \left( \x_f - \frac{t_f-t'}{t_f-t_i} (\x_f - \x_i) )  ,t' \right) \right) \right]_{ij} \, , \label{eq:U-def} 
\end{equation}
which, when ${\bf x}_f - {\bf x}_i = {\bf v} t$ only differs from~\eqref{eq:W-op-def} by an overall phase after identifying ${\bf x}$ with ${\bf x}_f$. Note that, as per the discussion above Eq.~\eqref{eq:W-op-def-expli}, the time evolution of $\Op$ in this expression is determined by $H_\varphi$.

Explicitly, the aforementioned structural property follows from
\begin{align}
    \langle W^{\vg,t}_{il,jk} \rangle (\Lb) &= \frac{1}{Z} {\rm Tr}_{\Hphi} \! \left\{ U_{kl}[(0,\Lb),(t,\vg t + \Lb)] U_{ij}[(t,\vg t),(0,{\bf 0})] e^{-\beta H_{\varphi}}  \right\} \nonumber \\
    &= \frac{1}{Z} {\rm Tr}_{\Hphi} \! \left\{ U_{ji}[(0,{\bf 0}),(-t,-\vg t)] U_{lk}[(-t,-\vg t -\Lb),(0,-\Lb)]  e^{-\beta H_{\varphi}}  \right\} \nonumber \\
    &= \frac{1}{Z} {\rm Tr}_{\Hphi} \! \left\{ U_{lk}[(-i\beta-t,-\vg t -\Lb),(-i\beta,-\Lb)]  U_{ji}[(0,{\bf 0}),(-t,-\vg t)] e^{-\beta H_{\varphi}}  \right\} \nonumber \\
    &= \frac{1}{Z} {\rm Tr}_{\Hphi} \! \left\{ U_{lk}[(-i\beta, -\Lb),(t-i\beta,\vg t -\Lb)]  U_{ji}[(t,\vg t),(0,{\bf 0} )] e^{-\beta H_{\varphi}}  \right\} \nonumber \\
    &\approx \frac{1}{Z} {\rm Tr}_{\Hphi} \! \left\{ U_{lk}[(0,i\beta \vg -\Lb),(t,\vg t -\Lb + i\beta \vg)]  U_{ji}[(t,\vg t),(0,{\bf 0} )] e^{-\beta H_{\varphi}}  \right\} \nonumber \\
    &= \langle W^{\vg,t}_{jk,il} \rangle (-{\bf L} + i {\bf v}/T ) \, . \label{eq:W-KMS}
\end{align}
The first line follows from using the definition of $\langle W^{\vg,t}_{il,jk} \rangle (\Lb)$ and writing it in terms of $U_{ij}$ as given by Eq.~\eqref{eq:U-def}. The second line follows from applying parity and time reversal operations, and relies on these operations being symmetries of the full theory defined by~\eqref{eq:Hamiltonian}. Note that due to the anti-unitary nature of the time reversal operator, the operator ordering gets inverted at this point. The third line follows from commuting the line operator with indices $lk$ with the thermal density matrix, and using that this is equivalent to evolving the operator along imaginary time by identifying $-i\beta$ as the time of propagation. The fourth line follows from performing a time translation by $t$ and a spatial translation by $\vg t$. The fifth line features an approximate sign because it relies on a long-time limit, analogous to the long-time limit that defines the Markovian regime~\eqref{eq:markovian}. 
It follows as a consequence of the fact that a spacetime translation of the line operator
\begin{equation}
    U_{lk}[(t_0, {\bf L} ),(t + t_0, {\bf v} t + {\bf L} )] = e^{i H_\varphi t_0 - i \Pphi \cdot \Lb } U_{lk}[(0, {\bf 0} ),(t , {\bf v} t  ) ] e^{-i H_\varphi t_0 + i \Pphi \cdot \Lb} \, ,
\end{equation}
is an analytic function of $t_0$ and the components of ${\bf L}$, and because of the fact that in the limit of an infinitely long line, this operator possesses a reparametrization invariance for real $t_0$, ${\bf L}$. Namely, the operator only depends on the combination ${\bf L} - {\bf v} t_0$, and is independent of $t_0 + \vg \cdot \Lb$ due to the fact that the line $(t' + t_0,\vg t' + \Lb)$ with $t' \in (-\infty, \infty)$ can be equivalently written as $(t',\vg t' - \vg t_0 + \Lb)$ by shifting the parameter $t'$. Because the operator is an analytic function of $t_0$ and the components of ${\bf L}$, it follows that this property --- i.e., that the operator only depends on ${\bf L} - {\bf v} t_0$ --- gets extended into the complex plane (the analytic continuation of a constant function on the real axis is a constant everywhere on the complex plane), and so the imaginary time shift by $-i\beta$ in the temporal argument can be moved into the position argument as a shift by $i\beta \vg$. 

Because of the explicit finite (but large) $t$ in all of our expressions, the fifth line in the above equation is approximate. In the formal limit $t\to \infty$,
\begin{equation}
    U_{lk}^\vg(t_0, \Lb) \equiv \lim_{t \to \infty} U_{lk}[(-t + t_0,-\vg t + \Lb),(t + t_0,\vg t + \Lb)] \, ,
\end{equation}
we have the equality
\begin{equation}
    U_{lk}^\vg(t_0, \Lb) = U_{lk}^\vg(0, \Lb - \vg t_0) \, , \label{eq:reparam}
\end{equation}
and therefore
\begin{equation}
    e^{\beta H_\varphi} U_{lk}^\vg(t_0, \Lb) e^{-\beta H_\varphi} = U_{lk}^\vg(t_0, \Lb + i \beta \vg ) \, ,
\end{equation}
which is the exact version of the relation we used in the derivation~\eqref{eq:W-KMS}. This is, however, impractical because the strict limit $t\to \infty$ leaves no information in Eq.~\eqref{eq:W-exp-S}. The more practical statement to make is that this KMS relation is a property of the $t$-extensive contribution in the exponent of Eq.~\eqref{eq:W-exp-S}, as the extent to which $U_{lk}$ is not invariant under reparametrizations as in~\eqref{eq:reparam} is controlled by the size of the shifts --- not by the total length of the lines, which is what defines $S$.

The KMS relation established by Eq.~\eqref{eq:W-KMS} provides the equilibration condition for the dynamics of the probe particle. As we shall see later in this section, this is analogous to the way in which the KMS relation for correlation functions of local operators provides the fluctuation-dissipation theorem.

\subsection{Relativistic environments in the rest frame of the probe particle} \label{sec:relativistic}

Because Lorentz transformations are a symmetry of the fundamental interactions that give rise to matter all around us, it is interesting to consider how the above considerations get reformulated if one calculates the time evolution in terms of quantities that are naturally defined in the rest frame of the particle.

To get an expression in terms of quantities in the rest frame of the particle, one can insert in between every operator in the trace unitary operators $U(\Lambda) U^\dagger(\Lambda)$ implementing a Lorentz transformation that maps the line $(t,\vg t)$ to $(t,{\bf 0})$. One obtains that the transformed object is
\begin{equation}
    \langle \tilde{W}^{\vg,t}_{il,jk} \rangle (\Lb') = \frac{1}{Z} {\rm Tr}_{\Hphi} \! \left\{ U_{kl}[(0,\Lb'),(t, \Lb')] U_{ij}[(t,{\bf 0}),(0,{\bf 0})] e^{-\beta \gamma H_{\varphi} - \beta \gamma \vg \cdot \Pphi }  \right\} \, ,
\end{equation}
where $\Lb'$ is the spatial separation between the two line operators in the new frame, and $\gamma = (1-|\vg|^2/c^2)^{-1}$ is the Lorentz boost factor. The density matrix of the environment describes an ensemble of states that favors states with momentum along $-\vg$, exactly as it results from boosting the thermal ensemble. Since we want to investigate the late time properties of this object, we have neglected the fact that a Lorentz transformation of the pair of lines which start at the same time in the original frame will not give a pair of lines that start at the same time in the new frame, because the difference between them does not depend on the large time $t$.

Following a similar sequence of steps as in Eq.~\eqref{eq:W-KMS}, we find that
\begin{align}
    \langle \tilde{W}^{\vg,t}_{il,jk} \rangle (\Lb') &= \frac{1}{Z} {\rm Tr}_{\Hphi} \! \big\{ U_{kl}[(0,\Lb'),(t, \Lb')] U_{ij}[(t,{\bf 0}),(0,{\bf 0})] e^{-\beta \gamma H_{\varphi} - \beta \gamma \vg \cdot \Pphi }  \big\} \nonumber \\
    &= \frac{1}{Z} {\rm Tr}_{\Hphi} \! \big\{ U_{ji}[(0,{\bf 0}),(-t,{\bf 0})] U_{lk}[(-t,-\Lb'),(0,-\Lb')]  e^{-\beta \gamma H_{\varphi} - \beta \gamma \vg \cdot \Pphi}  \big\} \nonumber \\
    &= \frac{1}{Z} {\rm Tr}_{\Hphi} \! \big\{ U_{lk}[(-i\gamma \beta-t, i \gamma \beta \vg  -\Lb'),(-i \gamma \beta,i \gamma \beta \vg-\Lb')] \nonumber \\ & \quad\quad\quad\quad\quad\quad\quad\quad\quad\quad\quad\quad\quad\quad \times  U_{ji}[(0,{\bf 0}),(-t, {\bf 0} )] e^{-\beta \gamma H_{\varphi} - \beta \gamma \vg \cdot \Pphi}  \big\} \nonumber \\
    &= \frac{1}{Z} {\rm Tr}_{\Hphi} \! \big\{ U_{lk}[(-i\gamma \beta, i \gamma \beta \vg  -\Lb'),(t-i \gamma \beta,i \gamma \beta \vg-\Lb')]  \nonumber \\ & \quad\quad\quad\quad\quad\quad\quad\quad\quad\quad\quad\quad\quad\quad \times U_{ji}[(t,{\bf 0}),(0, {\bf 0} )] e^{-\beta \gamma H_{\varphi} - \beta \gamma \vg \cdot \Pphi}  \big\} \nonumber \\
    &\approx \frac{1}{Z} {\rm Tr}_{\Hphi} \! \big\{ U_{lk}[(0, i \gamma \beta \vg  -\Lb'),(t,i \gamma \beta \vg-\Lb')]  U_{ji}[(t,{\bf 0}),(0, {\bf 0} )] e^{-\beta \gamma H_{\varphi} - \beta \gamma \vg \cdot \Pphi}  \big\} \nonumber \\
    &= \langle \tilde{W}^{\vg,t}_{jk,il} \rangle (-{\bf L}' + i \gamma {\bf v}/T ) \, , \label{eq:W-KMS-boosted}
\end{align}
where, in the logical sequence, a noteworthy difference from the previous derivation is that from the fourth to the fifth line we only used reparametrization invariance along the temporal direction.
One can readily see that the result of these manipulations is in fact equivalent to Eq.~\eqref{eq:W-KMS}, because the Lorentz boost relates $\Lb$ and $\Lb'$ via ${\bf L}' \cdot \vg = \gamma \Lb \cdot \vg$, and $\Lb' - (\Lb' \cdot \hat{\vg}) \hat{\vg} = \Lb - (\Lb \cdot \hat{\vg}) \hat{\vg} $, where $\hat{\vg}$ is the unit vector along the direction of motion of the probe particle.

Most importantly, this derivation makes transparent a fact that is not immediately apparent from Eq.~\eqref{eq:W-KMS}: it shows that the imaginary displacement by $i\gamma {\bf v}/T $ in the separation between the two line operators is a direct consequence of the fact that the states in the environment ensemble are prepared with a nonzero average momentum set by $e^{- \beta \gamma \vg \cdot \Pphi}$. This initial state generates the translation by $i\gamma {\bf v}/T$ when commuted with the line operators.

\subsection{KMS condition as an equilibrium condition of the master equation}

To make the connection with the properties of $S_{il,jk}(\Lb,\vg)$ explicit, it is convenient to relabel the indices as $a=(il)$, $b=(jk)$, and decompose the resulting matrix in terms of eigenvalues and eigenvectors. That is to say,
\begin{equation}
    S_{ab}(\Lb,\vg) = \sum_n v^{(R)}_{n,a}(\Lb,\vg) \lambda_n(\Lb,\vg) v^{(L)}_{n,b}(\Lb,\vg) \, , \label{eq:spectral-SofL}
\end{equation}
where $v^{(L)}_n$ and $v^{(R)}_n$ are left and right eigenvectors of $S_{ab}$ corresponding to the eigenvalue $\lambda_n$, at fixed $(\Lb,\vg)$, normalized such that the orthogonality condition $\sum_a v^{(L)}_{n,a} v^{(R)}_{m,a} = \delta_{nm}$ holds. Note that the generator of time evolution $S_{il,jk}(\Lb,\vg)$ need not be a Hermitian (self-adjoint) operator, and therefore its left and right eigenvectors do not necessarily agree. In fact, as one can prove rather straightforwardly, $\left( \langle W_{il,jk}^{\vg,t} \rangle (\Lb) \right)^* = \langle W_{li,kj}^{\vg,t} \rangle (-\Lb)$. The symmetry property it does possess (instead of hermiticity) is apparent from the discussion that follows.

Then, omitting non-extensive terms in $t$, i.e., in the Markovian limit, the KMS relation~\eqref{eq:W-KMS} implies that the evolution loop satisfies
\begin{align}
    \langle W_{ab}^{\vg,t} \rangle (\Lb) &= \sum_n v^{(R)}_{n,a}(\Lb,\vg) \exp\left( - t\lambda_n(\Lb,\vg)  \right)  v^{(L)}_{n,b}(\Lb,\vg) \nonumber \\
    &= \sum_n v^{(R)}_{n,b}(-\Lb + i \vg/T,\vg) \exp\left( - t\lambda_n(-\Lb + i \vg/T,\vg)  \right)  v^{(L)}_{n,a}(-\Lb + i\vg/T,\vg) \nonumber \\
    &= \langle W_{ba}^{\vg,t} \rangle (-\Lb+i\vg/T)
    \, ,
\end{align}
which, using the orthogonality condition and assuming that both left and right eigenvectors are a complete basis for the states in the internal degrees of freedom of the probe particle, implies that 
\begin{equation}
    \lambda_n(\Lb,\vg) = \lambda_n(-\Lb + i \vg/T,\vg) \, ,
\end{equation}
and (up to an arbitrary prefactor in the definition of the eigenvectors)
\begin{align}
    v^{(R)}_{n,a}(\Lb,\vg) = v^{(L)}_{n,a}(-\Lb + i\vg/T,\vg) \, , & & v^{(L)}_{n,b}(\Lb,\vg) = v^{(R)}_{n,b}(-\Lb + i \vg/T,\vg) \, .
\end{align}

All of the above amounts to saying, in a rather detailed fashion, that
\begin{equation}
    S_{il,jk}(\Lb,\vg) = S_{jk,il}(-\Lb + i \vg /T,\vg) \, , \label{eq:KMS-S}
\end{equation}
which is to say, the KMS relation in the correlation function of line operators gets directly imprinted onto the generator of time evolution of the reduced density matrix in the master equation~\eqref{eq:master-final}.

This has an immediate consequence regarding equilibration of the probe particle at late times. Coming back to Eq.~\eqref{eq:master-final}, we see that equilibrium solutions for the reduced density matrix $\tilde{\rho}$ satisfy
\begin{equation}
    S_{il,jk}(-i \vec{\partial}_\p , \vg(\p) ) \tilde{\rho}_{jk}^{\rm eq}(\p, \Delta \kb) = 0 \, , \label{eq:equil-cond}
\end{equation}
where we have included an arrow above the derivative to emphasize the direction in which it acts. As discussed when we introduced it, the order of the arguments is deliberate and consequential, emphasizing that the derivative also acts on the second argument of the function.

Viewed as a differential operator, $S_{il,jk}(-i \partial_\p , \vg(\p))$ also has a spectral decomposition akin (but different) to~\eqref{eq:spectral-SofL}, where the left and right eigenvectors are functions of $\p$ and the discrete index $a=(il)$, and $\Lb$ and $\vg$ no longer appear. We may write it as
\begin{equation}
    S_{ab}(-i \partial_\p , \vg(\p) ) = \sum_n w^{(R)}_{n,a}(\p) \sigma_n w^{(L)}_{n,b}(\p) \, , \label{eq:spectral-Sofp}
\end{equation}
where we have denoted the eigenvalues by $\sigma_n$, and we normalize the left and right eigenvectors $w^{(L)}$ and $w^{(R)}$ as 
\begin{equation}
    \sum_a \int_\p  w^{(L)}_{n,a}(\p) w^{(R)}_{m,a}(\p) = \delta_{nm} \, .
\end{equation}

While this may (again) seem like a highly formal expression, in fact it contains information we can exploit. First, note that unitarity implies that 
\begin{equation}
    \sum_{i} S_{ii,jk}({\bf 0},\vg) = 0 \, , \label{eq:unitarity}
\end{equation}
because at zero spatial separation the line operators coming from the amplitude and its complex conjugate exactly compensate each other, for any initial density matrix in~\eqref{eq:evloution-result-inbin}. Therefore, the constant matrix
\begin{equation}
    w_{0,(il)}^{(L)}(\p) = \delta_{il} \, ,
\end{equation}
where $\delta_{il}$ is a Kronecker delta, is a left eigenvector of $S$ with eigenvalue $\sigma_0 = 0$. We have assigned the label $n=0$ to it. This means that there is a corresponding right eigenvector $w_{0,(jk)}^{(R)}(\p)$ with zero eigenvalue, i.e., that satisfies Eq.~\eqref{eq:equil-cond}, and is therefore an equilibrium solution for the dynamics of $\tilde{\rho}$.

What is this equilibrium state? This is where the KMS relation~\eqref{eq:KMS-S} comes into play. Concretely, the KMS relation coupled with unitarity~\eqref{eq:unitarity} implies that
\begin{equation}
    \sum_{i} S_{jk,ii}(i \vg /T,\vg) = 0 \, , \label{eq:KMS-plus-unitarity}
\end{equation}
which suggests to look for an equilibrium state of the form $\tilde{\rho}^{\rm eq}_{jk} = \delta_{jk} f_{\rm eq}(\p) $. That is to say, we seek solutions to
\begin{equation}
    S_{il,jj}(-i \vec{\partial}_\p , \vg(\p) ) f_{\rm eq}(\p) = 0 \, . \label{eq:equil-cond-f}
\end{equation}

In general, solutions to this equation may be complicated. However, a clear structure emerges in the limit in which the characteristic mass/energy/momentum scale of the probe's dispersion relation, which we denote by $\Lambda$, is much larger than the rest of the scales in the problem (the internal scales of the environment and its temperature). In these cases, we may write $E(\p ) = \Lambda \, \varepsilon (\p/\Lambda)$ where $\varepsilon({\bf u})$ does not depend on $\Lambda$\footnote{Relativistic ($E(\p) = \sqrt{\p^2 + M^2}$) and non-relativistic heavy particles ($E(\p) = \p^2/(2M)$) exemplify this. By identifying $\Lambda = M$, one obtains $\varepsilon({\bf u}) = \sqrt{1 + {\bf u}^2}$ and $\varepsilon({\bf u}) = {\bf u}^2/2$, respectively.}. Crucially, this means that the velocity components of the probe $[\vg]_i \equiv v_i$ satisfy
\begin{align}
    v_i(\p) = \frac{\partial E}{\partial p_i} = (\partial_i \varepsilon) (\p/\Lambda) \, ,& & {\rm and} & & \partial_i v_j = \frac{1}{\Lambda} (\partial_i \partial_j \varepsilon) (\p/\Lambda) \, .
\end{align}
This means that momentum derivatives of the velocity are suppressed by powers of $\Lambda^{-1}$. Thus, in the limit where $\Lambda$ is much larger than the scales in $S_{il,jj}$, we may ignore the action of derivatives on the second argument and only keep their action on $f_{\rm eq}$, up to the aforementioned power corrections. Then, in light of Eq.~\eqref{eq:KMS-plus-unitarity}, the solution to Eq.~\eqref{eq:equil-cond-f} at leading order in $\Lambda^{-1}$ is the eigenfunction of the operator $-i \partial_\p$ with ``eigenvalue'' $i \vg(\p)/T$, i.e.,
\begin{align}
    -i \partial_\p f_{\rm eq} = i \frac{\vg(\p)}{T} f_{\rm eq} & & \implies & & f_{\rm eq} = f_0 \exp \left( - \frac{E(\p)}{T} \right) \, .
\end{align}

In conclusion, all of this shows that in the limit where the characteristic momentum scale of the probe is well separated from those of the medium, there is an equilibrium solution to the master equation~\eqref{eq:master-final}, given by
\begin{equation}
    \tilde{\rho}^{\rm eq}_{jk} = \frac{1}{\tilde{Z}} \delta_{jk} \exp \left( - \frac{E(\p)}{T} \right) \, , \label{eq:equil-solution}
\end{equation}
with $\tilde{Z}$ setting the normalization.
This solution is exactly what one would have expected from general grounds in statistical mechanics: in the limit where the medium does not modify the dispersion relation of the probe particle in any substantial way (here enforced by the separation between $\Lambda$ and other scales), the equilibrium solution is given by thermal weights for each energy eigenstate as prescribed by the Boltzmann distribution.

Note that the equilibrium condition does not make any reference to the value of $\Delta \kb$. This means that, as far as the dynamics generated by Eq.~\eqref{eq:master-final} goes, Eq.~\eqref{eq:equil-solution} holds with $\tilde{Z}$ an arbitrary function of $\Delta \kb$. However, in the derivation of Eq.~\eqref{eq:master-final} we assumed that the support of $\tilde{\rho}$ as a function of $\Delta \kb$ must be localized around small values thereof. In general, for $|\Delta \kb| > 0$, corrections to the dynamics due to the terms we neglected in the expansion of $E(\p + \kb)$ will affect the equilibrium solution. These corrections can be large for entries of $\tilde{\rho}$ with $|\Delta \kb| \sim \Lambda$, consistent with the expectation that the equilibrium solution should be localized around the diagonal in momentum. 

On the other hand, if there were no large scale separation between $\Lambda$ and the other scales in the problem, one would have to return to Eq.~\eqref{eq:equil-cond-f} and study it in full generality. These situations correspond to scenarios in which the interaction with the environment can strongly modify the free dispersion relation of the probe particle --- in contrast with the previous case where the large separation between $\Lambda$ and the rest of scales guarantees that such corrections remain small.

\section{Stochastic dynamics of the probe particle}
\label{sec:stochastic}

The fact that the quantum master equation~\eqref{eq:master-final} does not couple different $\Delta \kb$ sectors, together with the fact that its derivation relies on taking $|\Delta \kb|$ to be small, is highly suggestive of a stochastic description of the motion of the probe particle, where one only keeps track of the net occupancies of each momentum state (i.e., of the diagonal entries of the reduced density matrix $\rho_P$). In fact, we already advanced the object that generates the dynamics of these occupancies: the momentum change probability $P(\kb;\vg,t)$ that we introduced in Eq.~\eqref{eq:prob-k-loss}. 
Following through the steps carried out in section~\ref{sec:quantum}, the natural descendant from Eq.~\eqref{eq:master-final} would be
\begin{equation}
    \partial_t f_i(\p;t) = - \sum_j S_{ii,jj}(-i \partial_\p,\vg(\p)) f_j(\p;t) \, , \label{eq:stochastic-conjecture}
\end{equation}
with no implicit summation over repeated indices, and where $f_i(\p)$ would be the diagonal elements of the probe particle density matrix $\rho_P$.

However, if one does not want to make uncontrolled approximations starting from the full quantum dynamics, there is a limitation to this conclusion. Namely, a general hermitian operator $\Op_{ij}$ need not preserve the symmetry under exchanging the internal labels $i,j$ that the free probe particle possesses. In general, we have
\begin{equation}
    \Op_{ij}(\x) = \sum_a T^a_{ij} \Op^a(\x) \, , \label{eq:O-decomp}
\end{equation}
where the $T^a$ matrices form a basis of Hermitian matrices, which can be taken to be the generators of $U(N)$ for $N$ the dimension of the internal label space. That is to say, $N$ is such that $i,j \in \{1,2,\ldots,N\}$. In general, the set of matrices $T^a$ in this decomposition with distinct, nonzero operator coefficients will not be simultaneously diagonalizable. Therefore, the evolution operator $S_{il,jk}$ of the quantum theory will generally couple diagonal and non-diagonal components in the labels $jk$ of the density matrix $\tilde{\rho}_{jk}$, and as such the dynamics of the diagonal elements in the quantum theory will not be reproduced by a classical equation such as~\eqref{eq:stochastic-conjecture} if one truncates the evolution equations to only take these diagonal elements as input and output.

Consequently, if one wants to write Eq.~\eqref{eq:stochastic-conjecture}, and faithfully reproduce the dynamics that the quantum master equation~\eqref{eq:master-final} generates, additional assumptions are needed. 
Two possibilities where equations of the same form as Eq.~\eqref{eq:stochastic-conjecture} follow are:
\begin{enumerate}
    \item The case where the full theory is symmetric under transformations among the indices $i,j$. That is to say, even if specific instances of the action of the operator $\Op_{ij}$ might break this symmetry, the dynamics of the environment does preserve said symmetry, and as such the thermal ensemble will do so as well. In this case, if one initializes the reduced density matrix proportional to $\delta_{jk}$ in the internal labels, the evolution will necessarily maintain the occupancies in the internal labels diagonal equally distributed. Therefore, in this case the evolution of the number of probe particles summed over internal labels $f(\p;t) \equiv \sum_i f_i(\p;t)$ is governed by
    \begin{equation}
        \partial_t f(\p;t) = - S_{\rm ave}(-i \partial_\p,\vg(\p)) f(\p;t) \, , \label{eq:stochastic-averaged}
    \end{equation}
    where $S_{\rm ave}$ is defined from the late-time limit of the initial and final-state averaged evolution loop, i.e., via
    \begin{equation}
        \partial_t \left[ \sum_{i,j} \langle W^{\vg,t}_{ii,jj} \rangle (\Lb) \right] = -\left[ S_{\rm ave}(\Lb,\vg) + O(t^{-1}) \right] \sum_{k,l} \langle W^{\vg,t}_{ll,kk} \rangle (\Lb) \, . \label{eq:markovian-averaged}
    \end{equation}
    \item The case where the operator $\Op_{ij}(\x)$ may be written as in Eq.~\eqref{eq:O-decomp} and the matrices $T^a$ appearing in the decomposition are simultaneously diagonalizable. In this case, we may simply treat the dynamics of the projections onto the eigenstates of these matrices separately. Without loss of generality, we may define the $i,j$ indices such that they describe mutually decoupled degrees of freedom. As such, each probe particle species evolves independently, with a separate evolution equation for each value of $i$
    \begin{equation}
        \partial_t f_i(\p;t) = -S_{ii,ii}(-i \partial_\p,\vg(\p)) f_i(\p;t) \, . \label{eq:stochastic-independent}
    \end{equation}
\end{enumerate}

Both of these equations~\eqref{eq:stochastic-averaged} and~\eqref{eq:stochastic-independent} are of the form
\begin{equation}
    \partial_t f(\p;t) = - S(-i \partial_\p,\vg(\p))) f(\p;t) \, , \label{eq:Kolmogorov}
\end{equation}
and define stochastic processes for the momentum of the probe particle, where $f$ is the probability density function of the probe particle's momentum at time $t$. More specifically, they are Kolmogorov equations describing the (non-Gaussian) momentum diffusion of the probe particle. Furthermore, their equilibrium state is $f_{\rm eq} \propto \exp \left( - E(\p)/T \right)$ by virtue of the KMS relation for $S$, given in Eq.~\eqref{eq:KMS-S}.

In this way, Eq.~\eqref{eq:Kolmogorov} defines the Langevin equation~\eqref{eq:Langevin}, where the stochastic force $\xi$ is determined by the momentum transfer probability $P({\bf k};{\bf v},t)$ given by~\eqref{eq:P-of-W-averaged}.\footnote{As written, this only applies to situation 1 of the paragraph above. In situation 2, this probability is extracted from a projection of the evolution loop instead of an average.}

\subsection{Equilibration condition, KMS relation, detailed balance, and a tower of fluctuation-dissipation relations}

The KMS condition in terms of line operators~\eqref{eq:W-KMS} also has a natural interpretation in terms of a property that the transition probability $P$ satisfies in a stochastic process around an equilibrium solution: detailed balance. Indeed, from Eq.~\eqref{eq:P-of-W-averaged}
we see that upon using~\eqref{eq:W-KMS} one obtains
\begin{equation}
    P({\bf k};{\bf v},t) = e^{{\bf k} \cdot {\bf v}/T } P(-{\bf k};{\bf v},t) \, , \label{eq:detailed-balance}
\end{equation}
which is exactly the detailed balance condition
\begin{align}
    \exp \left(-\frac{E({\bf p})}{T} \right) P({\bf k};{\bf v}({\bf p}) ) = \exp \left(-\frac{E({\bf p} - {\bf k})}{T} \right) P(-{\bf k};{\bf v}({\bf p} - {\bf k}) ) \, ,
\end{align}
to first order in an expansion at small $\kb$ around the reference momentum $\p$ --- simply following from the fact that $E(\p + \kb) \approx E(\p) + \kb \cdot \vg $ and using the assumption that $\vg(\p+\kb) \approx \vg(\p)$ for the momentum changes $\kb$ in which $P$ has support [which can be traced back to the assumption we made to derive the equilibrium solution~\eqref{eq:equil-solution}, namely, the existence of a large scale $\Lambda$ that characterizes the dispersion relation as $E(\p) = \Lambda \, \varepsilon( \p/\Lambda )$].

This is to be contrasted with the standard argument by which Brownian dynamics leads a probe particle to reach thermal equilibrium with its environment: the fluctuation-dissipation theorem~\cite{Kubo1966}. In its simplest form for our present purposes, it is given by the Einstein relation between the drag force ${\bf F}_D$ and the momentum diffusion coefficient $\kappa$
\begin{equation}
    \vg \kappa = 2 T {\bf F}_D \, , \label{eq:Fluct-Diss}
\end{equation}
which, on the one hand, is a consistency condition for the Langevin equation~\eqref{eq:Langevin} if the stochastic force is Gaussian, and, on the other hand, a consequence of statistical physics considerations in the case where fluctuations of the stochastic force are not correlated with the velocity ${\bf v}$ (see, e.g., sections 4 and 7 of~\cite{Kubo1966}). For example, a non-relativistic heavy particle coupled to a relativistic environment will experience a random force where the ``kicks'' it receives will be independent of the (small) velocity of the particle. 

In our setup, no assumption is made regarding the dependence of the stochastic force on the velocity of the probe particle. Indeed, away from the non-relativistic limit (or, more generally, as soon as the stochastic force exhibits a quantitatively important dependence on the velocity), Eq.~\eqref{eq:Fluct-Diss} is no longer valid, and it is instead replaced by Eq.~\eqref{eq:KMS-plus-unitarity}. The latter equation is, in the same sense of~\cite{Rajagopal:2025rxr} for the case of heavy quarks in gauge theories, a generalization of the Einstein relation to the generic case where the statistics of the momentum transfer to the particle is non-Gaussian, applicable to the kinetic theory of a probe particle coupled to any quantum bath at finite temperature, regardless of the strength of their coupling and the internal dynamics of the bath.

\paragraph{A velocity-dependent family of Fluctuation-Dissipation relations.}

The KMS condition~\eqref{eq:W-KMS} underlying all of our results can be cast as a family of relations between correlators of the operator $\Op$ of different order. To see this, it is convenient to start from the counterpart of Eq.~\eqref{eq:KMS-plus-unitarity} at the level of the evolution loop:
\begin{equation} \label{eq:W-KMS-plus-unitarity}
    \sum_i \langle W_{jk,ii}^{\vg,t} \rangle (i\vg / T) = \delta_{jk} \, ,
\end{equation}
and consider the series expansion of the evolution loop (summed over the internal index $i$) along the spatial separation $L_\parallel = \hat{\vg} \cdot \Lb$ parallel to $\vg$ (where $\hat{\vg} = \vg / |\vg|$), keeping the separation along the orthogonal direction fixed to zero
\begin{equation}
    \sum_i \langle W^{\vg,t}_{jk,ii} \rangle (\Lb = L_\parallel \hat{\vg} ) = \sum_{m=0}^{\infty} c^{(m)}_{jk}(|\vg|) \frac{L_\parallel^m}{m!} \, , 
\end{equation}
which, after imposing~\eqref{eq:W-KMS-plus-unitarity}, reveals a sum rule
\begin{equation}
    \sum_{m=1}^{\infty} c^{(m)}_{jk}(|\vg|) \frac{(i |\vg|/T )^m}{m!} = 0 \label{eq:v-dependent-sum-rule}
\end{equation}
between the coefficients of the series. Here we have used that $c^{(0)}_{jk} = \delta_{jk}$. This is a family of sum rules valid, separately, for each value of $|\vg|$.

The family of fluctuation-dissipation relations between correlation functions that we anticipated becomes apparent when we note that the coefficients $c^{(m)}_{jk}(|\vg|)$ are exactly such quantities. Concretely, these expansion coefficients are nothing more than correlation functions of $\Op$ inserted along the line operators that define the evolution loop. This is most apparent from Eq.~\eqref{eq:W-op-def}; explicitly,
\begin{align}
    &\frac{\partial}{\partial \x} \left[ {\rm P} \exp \left( -i \int_{t_0}^t dt' \mathcal{O}(\x + t' \vg  ,t') \right) \right]_{ij} \nonumber \\
    &= -i \int_{t_0}^t ds \left[ {\rm P} \exp \left( -i \int_{s}^t dt' \mathcal{O}(\x +t' \vg  ,t') \right) \right]_{ik} [\partial_{\x} \mathcal{O}(\x +s \vg  ,s)]_{kl}  \nonumber \\ & \quad \quad \quad \quad \times \left[ {\rm P} \exp \left( -i \int_{t_0}^s dt' \mathcal{O}(\x + t' \vg  ,t') \right) \right]_{lj} \, , \label{eq:W-op-Lb-deri}
\end{align}
i.e., the same line operator as in Eq.~\eqref{eq:W-op-def}, but with an explicit insertion of the operator $\partial_\x \Op$, summed over the line. This can be written more succinctly as
\begin{equation}
    \frac{\partial}{\partial \x} W^{\vg,\x}_{ij}(t,t_0) = - i \int_{t_0}^t ds \, W^{\vg,\x}_{ik}(t,s) [\partial_\x \mathcal{O}(\x + s \vg  ,s)]_{kl}  W^{\vg,\x}_{lj}(s,t_0) \, .
\end{equation}
Higher order derivatives of the line operator generate more insertions of $\partial_\x \Op$ or higher derivatives thereof. To be explicit, a few low-order coefficients are
\begin{align}
    c_{jk}^{(1)} &= \frac{i}{Z} \int_{0}^t ds \, {\rm Tr}_{\mathcal{H}_\varphi } \! \Big[ W^{\vg,{\bf 0} }_{ji}(t,0) e^{-\beta H_\varphi } W^{\vg,{\bf 0}}_{ii'}(0,s)  [\hat{\vg} \cdot  \partial_\x \mathcal{O}( s \vg  ,s)]_{i'k'} W^{\vg,{\bf 0}}_{k'k}(s,t) \Big] \, , \label{eq:cjk1} \\
    c_{jk}^{(2)} &= \frac{1}{Z} \int_{0}^t ds' ds \, {\rm Tr}_{\mathcal{H}_\varphi } \! \Big[ W^{\vg,{\bf 0} }_{jj'}(t,s') [\hat{\vg} \cdot  \partial_\x \mathcal{O}( s' \vg  ,s')]_{j'l'} W^{\vg,{\bf 0} }_{l'i}(s',0) e^{-\beta H_\varphi } \nonumber \\ & \quad\quad\quad\quad\quad\quad\quad\quad\quad\quad\quad\quad\quad\quad\,\,\, W^{\vg,{\bf 0}}_{ii'}(0,s)  [\hat{\vg} \cdot  \partial_\x \mathcal{O}( s \vg  ,s)]_{i'k'} W^{\vg,{\bf 0}}_{k'k}(s,t) \Big] \, ,  \label{eq:cjk2} \\
    c_{jk}^{(3)} &= \frac{1}{Z} \int_{0}^t ds' ds \, {\rm Tr}_{\mathcal{H}_\varphi } \! \Big[ W^{\vg,{\bf 0} }_{jj'}(t,s') [\hat{\vg} \cdot  \partial_\x \mathcal{O}( s' \vg  ,s')]_{j'l'} W^{\vg,{\bf 0} }_{l'i}(s',0) e^{-\beta H_\varphi } \nonumber \\ & \quad\quad\quad\quad\quad\quad\quad\quad\quad\quad\quad\quad\quad\,\, W^{\vg,{\bf 0}}_{ii'}(0,s)  [(\hat{\vg} \cdot  \partial_\x)^2 \mathcal{O}( s \vg  ,s)]_{i'k'} W^{\vg,{\bf 0}}_{k'k}(s,t) \Big] \nonumber \\
    & + \frac{2i}{Z} \int_{0}^t  ds' ds \int_0^{s} ds'' \, {\rm Tr}_{\mathcal{H}_\varphi } \! \Big[ W^{\vg,{\bf 0} }_{jj'}(t,s') [\hat{\vg} \cdot  \partial_\x \mathcal{O}( s' \vg  ,s')]_{j'l'} W^{\vg,{\bf 0} }_{l'i}(s',0) e^{-\beta H_\varphi } \nonumber \\ & \quad\quad\quad\quad W^{\vg,{\bf 0}}_{ii'}(0,s'')  [\hat{\vg} \cdot  \partial_\x \mathcal{O}( s'' \vg  ,s'')]_{i'h'} W^{\vg,{\bf 0}}_{h'g'}(s'',s)  [\hat{\vg} \cdot  \partial_\x \mathcal{O}( s \vg  ,s)]_{g'k'} W^{\vg,{\bf 0}}_{k'k}(s,t) \Big] \, ,   \label{eq:cjk3}
\end{align}
which encode the drag force, the variance of the momentum fluctuations, and the skewness of the momentum transfer that the probe particle experiences --- as one may see directly from Eq.~\eqref{eq:P-of-W-averaged}, where it is clear that derivatives the with respect to ${\bf L}$ of the evolution loop are exactly the moments of the momentum change probability. For the last two coefficients in the above equations (\eqref{eq:cjk2} and~\eqref{eq:cjk3}) we made use of translational invariance so that each line operator was acted upon by a derivative at least once, which simplifies some of the algebra.

It is comforting to see that Eq.~\eqref{eq:Fluct-Diss} emerges from the sum rule~\eqref{eq:v-dependent-sum-rule} in the zero velocity limit: if one considers the series expansion of the coefficients $c_{jk}^{(m)}$ as a function of $|\vg|$ and uses~\eqref{eq:v-dependent-sum-rule} to obtain relations between the coefficients of each power of $|\vg|$, it follows that
\begin{equation}
    i\frac{ \partial_{|\vg|} c_{jk}^{(1)}(0)}{T}  - \frac{c_{jk}^{(2)}(0)}{2T^2} = 0 \, . \label{eq:NR-constraint}
\end{equation}
Then, recognizing that
\begin{align}
    c_{jj}^{(1)} &= \left[ \partial_{L_\parallel} \sum_{i,j} \langle W^{\vg,t}_{jj,ii} \rangle (\Lb = L_\parallel \hat{\vg} )  \right]_{L_\parallel = 0} = i \int_{\kb}  (\hat{\vg} \cdot \kb) \,  P(\kb;\vg,t) \equiv i t \hat{\vg} \cdot {\bf F}_D \, , \\
    c_{jj}^{(2)} &= \left[ \partial_{L_\parallel}^2 \sum_{i,j} \langle W^{\vg,t}_{jj,ii} \rangle (\Lb = L_\parallel \hat{\vg} )  \right]_{L_\parallel = 0} = - \int_{\kb} (\hat{\vg} \cdot \kb)^2 \,  P(\kb;\vg,t) \equiv_{\vg \to {\bf 0}} - t \kappa \, ,
\end{align}
it follows from~\eqref{eq:NR-constraint} that
\begin{equation}
    \kappa = 2 T \, [\partial_{|\vg|} (\hat{\vg} \cdot {\bf F}_D)]_{|\vg| = 0 } \, ,
\end{equation}
which, provided that ${\bf F}_D$ is parallel to $\vg$, is equivalent to the Einstein relation~\eqref{eq:Fluct-Diss}.

We emphasize that, for $|\vg| > 0$, this relation is not guaranteed, and explicit calculations (such as the one by Gubser~\cite{Gubser:2006nz} that we mentioned in the introduction) typically break it. For a general velocity, only the sum rule~\eqref{eq:v-dependent-sum-rule} is guaranteed to hold.

\paragraph{Fluctuation-Dissipation relations in the static limit.}

Furthermore, the KMS condition implies a tower of fluctuation-dissipation relations in the static limit $\vg = {\bf 0}$. To see this, first perform a series expansion of the evolution loop in both ${\bf v}$ and ${\bf L}$. Assuming rotational symmetry, the evolution loop can be written in terms of the magnitude of the velocity $v \equiv |\vg|$, the longitudinal separation between the lines $L_\parallel = \Lb \cdot \vg/v$, and the transverse separation $\Lb_\perp = \Lb - L_\parallel \vg/v$, which by the same token can be taken to be any one of the directions perpendicular to the velocity. The evolution loop may then be expanded as
\begin{equation}
    \langle W^{\vg,t}_{il,jk} \rangle (\Lb) = \sum_{n,m,p=0}^{\infty} c^{(n,m,p)}_{ij,kl} \frac{v^n}{n!} \frac{L_\parallel^m}{m!} \frac{L_\perp^{2p}}{(2p)!} \, , 
\end{equation}
where we have used that the dependence on $L_\perp$ has to be through $L_\perp^2$, by symmetry. Then, the KMS condition for line operators~\eqref{eq:W-KMS} implies that
\begin{align}
    \sum_{n,m,p=0}^{\infty} c^{(n,m,p)}_{ij,kl} \frac{v^n}{n!} \frac{L_\parallel^m}{m!} \frac{L_\perp^{2p}}{(2p)!} &= \sum_{n,m,p=0}^{\infty} c^{(n,m,p)}_{ij,kl} \frac{v^n}{n!} \frac{(-L_\parallel + i v/T)^m}{m!} \frac{L_\perp^{2p}}{(2p)!} \nonumber \\
    &= \sum_{n,m,p=0}^{\infty} \left[ \sum_{q=0}^n \binom{n}{q} (-1)^m (i/T)^q c^{(n-q,m+q,p)}_{ij,kl} \right] \frac{v^n}{n!} \frac{L_\parallel^m}{m!} \frac{L_\perp^{2p}}{(2p)!} \, ,
\end{align}
meaning that the expansion coefficients satisfy the sum rules
\begin{equation}
    c^{(n,m,p)}_{ij,kl} = \sum_{q=0}^n \binom{n}{q} (-1)^m (i/T)^q c^{(n-q,m+q,p)}_{ij,kl} \, . \label{eq:sum-rules}
\end{equation}
We note that the rule specified by $n=0$ is either trivial (even $m$) or simply yields a consequence of parity (odd $m$). Furthermore, the rules $(1,m,p)$ are equivalent to $(2,m-1,p)$ for all odd $m \geq 1$.

As before, these expansion coefficients are nothing more than correlation functions of derivatives of $\Op$ inserted along the line operators that define the evolution loop. The coefficients of the expansion with respect to the spatial separation can be obtained by taking derivatives as in Eq.~\eqref{eq:W-op-Lb-deri}, and those with respect to $v$ can be obtained via
\begin{align}
    &\frac{\partial}{\partial \vg} \left[ {\rm P} \exp \left( -i \int_{t_0}^t dt' \mathcal{O}(\Lb + t' \vg  ,t') \right) \right]_{ij} \nonumber \\
    &= -i \int_{t_0}^t ds \left[ {\rm P} \exp \left( -i \int_{s}^t dt' \mathcal{O}(\Lb +t' \vg  ,t') \right) \right]_{ik} s \, [\partial_{\Lb} \mathcal{O}(\Lb + s \vg  ,s)]_{kl}  \nonumber \\ & \quad \quad \quad\quad  \times \left[ {\rm P} \exp \left( -i \int_{t_0}^s dt' \mathcal{O}(\Lb + t' \vg  ,t') \right) \right]_{lj} \, , \label{eq:W-v-derivative}
\end{align}
with a slight practical difference in that this derivative necessarily acts on the two lines in the evolution loop~\eqref{eq:eff-evol-def}, whereas the derivative with respect to $\Lb$ can be taken to act on only one of the lines due to the overall shift symmetry.

As a final remark in this section, let us note that if $H_\varphi$ does not possess rotational symmetry one could still derive nontrivial fluctuation-dissipation relations, albeit the derivation would have to be suitably modified to handle the different spatial components of $\vg$ and $\Lb$ on a separate footing.

\subsection{When and how a Gaussian process emerges}

No matter how non-Gaussian the ``microscopic'' momentum transfer probability $P(\kb)$ is, the central limit theorem dictates that if one waits a sufficiently long time for stochastic kicks to accumulate, as long as they are sampled from the same underlying distribution, the sum of the kicks will (approximately) follow a Gaussian distribution, with mean and variances given by the sum of their microscopic counterparts.

In general, however --- as we have just seen --- the truncation of the dynamics in the Kolmogorov equation~\eqref{eq:Kolmogorov} to its Gaussian characteristics does \textit{not} lead to the correct late-time equilibrium solution. In fact, the only place where this truncation is always consistent is around the zero velocity limit. As such, this is the only regime where Gaussian dynamics can be an accurate description of the full stochastic dynamics --- because only here it approaches the correct equilibrium solution.
Indeed, the truncation of Eq.~\eqref{eq:Kolmogorov} to the Gaussian features of the momentum transfer probability $P(\kb;\vg)$ leads to the well-known Fokker-Planck equation
\begin{equation}
    \partial_t f(\p;t) = \partial_i \left( F_i f \right) + \frac12 \partial_i \partial_j \left( \kappa_{ij} f \right) \, ,
\end{equation}
and evaluating $F_i$, $\kappa_{ij}$ in the $\vg \to {\bf 0}$ limit (in which $\kappa_{ij} = \delta_{ij} \kappa$), by virtue of the Fluctuation-Dissipation theorem~\eqref{eq:Fluct-Diss}, yields an equilibrium distribution given by
\begin{equation}
    f_{\rm eq}^{\rm NR}({\bf p}) = \exp \left( - \frac{\p^2}{2MT} \right) \, ,
\end{equation}
i.e., the Boltzmann distribution for a dispersion relation of a non-relativistic particle $E(\p) = \p^2/(2M)$.

Conversely, this means that the non-Gaussian features of the momentum transfer probability are generally crucial for the Boltzmann distribution to be reached as a stationary state of the dynamics in regimes where the probe particle does not follow the non-relativistic particle dispersion relation. 
The fact that the kinematic range of applicability of the Gaussian truncation will increase for larger values of $M$ --- because the degree to which the stochastic kicks from the environment to the particle become closer to identically distributed increases --- is consistent with this in the sense that as the ratio $M/T$ increases, more and more of the probability of finding the particle gets localized around $\p = {\bf 0}$ for situations close to equilibrium, and as such the error incurred in using this approximation in terms of the net probability becomes smaller and smaller. That being said, unless the Gaussian features of the momentum transfer probability happen to satisfy the Einstein relation, its non-Gaussian features will be necessary to reach the correct equilibrium distribution.

\section{Outlook} \label{sec:outlook}

We have established a universal equilibration condition for probe particles propagating in a quantum thermal environment, following only from unitarity, having spacetime translations, parity and time-reversal as symmetries, and the probe particle limit encoded in the approximation made in going from Eq.~\eqref{eq:Hamiltonian} to~\eqref{eq:Hamiltonian-approx}. In addition to explicitly demonstrating that kinetic equilibration can be achieved, it gives a concrete answer for what quantity to calculate in order to quantify the properties of the equilibration process. This quantity is given by Eq.~\eqref{eq:eff-evol-def}, which we have referred to as the evolution loop. The fact that fast-moving probe particles preferentially lose, rather than gain energy, is clear from the equilibration condition~\eqref{eq:detailed-balance}, and can ultimately be traced back to the statistical average over the environment degrees of freedom.

With slight modifications regarding initial state preparation, the evolution loop has a long history in the context of gauge theories, where it is a specific instance of a broader class of observables known as Wilson loops. While the discovery of the equilibration condition for heavy quarks is rather recent~\cite{Rajagopal:2025rxr}, specific instances (or limits thereof) of that same Wilson loop had appeared long before in the context of heavy quark diffusion~\cite{Casalderrey-Solana:2006fio,Casalderrey-Solana:2007ahi} and jet quenching~\cite{Liu:2006ug,DEramo:2010wup,DEramo:2012uzl,Benzke:2012sz,Ghiglieri:2015ala}. The discovery of the equilibration condition for heavy quarks followed the realization that the velocity dependence of this Wilson loop at strong coupling in $\mathcal{N}=4$ SYM theory~\cite{Rajagopal:2025ukd} --- all the way from the non-relativistic limit $v=0$ to the ultra-relativistic limit $v \to c$ --- possessed structural properties that allowed for equilibration to take place. First steps toward phenomenological studies applying these results were taken in~\cite{Rajagopal:2026urq}. The complementary limit of weak coupling, along with an extension to a broad class of gauge theories, was examined in~\cite{DuPlessis:2026pyr}.
There has also been plenty of interest in developing an understanding of jets in heavy-ion collisions as open quantum systems~\cite{Mehtar-Tani:2024smp,Mehtar-Tani:2025xxd,Vaidya:2020cyi,Vaidya:2020lih,Vaidya:2021vxu,Vaidya:2026yfa}, where similar operators with Wilson lines play a central role.

While the mathematical machinery we used here is motivated from a quantum field theory perspective, the physics encoded in the evolution loop can also be presented in terms of different descriptions when there is an overlapping regime of applicability. For example, it is straightforward to cast Eq.~\eqref{eq:Kolmogorov} in the form of a Boltzmann kinetic equation
\begin{equation}
    \partial_t f(\p;t) = \int_{\kb} \left[ C(\kb;\p+\kb) f(\p+\kb;t) - C(\kb;\p) f(\p;t)  \right] \, ,
\end{equation}
where $C(\kb)$ is nothing more than the Fourier transform of $-S(\Lb)$ up to an overall constant; the subtraction term in the square brackets above is nothing more than this constant, consistent with our convention that $S({\bf 0};{\bf v}) = 0$. In turn, $C(\kb;\p)$ is nothing more than the scattering rate of the probe particle to go between different momentum states. This form is most useful when one knows, a priori, what the constituents of the environment are; in particular, if these constituents are quasi-particles. If so, these scattering rates can be written in terms of (a collection of) $n \leftrightarrow n'$ scattering processes and directly encode properties of the (quasi-)particles that form the medium. With further assumptions, the Boltzmann transport formalism then allows one to go further and make the environment itself dynamical --- a big advantage compared to our setup, where we always assume that the environment is described by a thermal density matrix. On the other hand, our approach does not assume a specific form of the constituents of the medium, nor whether they behave as quasi-particles. In this sense, the framework we have developed here specifies exactly what the generalization of $n \leftrightarrow n'$ scattering processes in a Boltzmann equation is in situations where the environment is strongly coupled and does not admit a quasi-particle description.

Another example where the evolution loop encodes well-known physics is some instances of classical radiative phenomena, where radiation is emitted by a probe particle. Recently, it was shown that if the velocity characterizing the evolution loop is taken to be larger than the speed of light in a medium described by macroscopic electrodynamics, this loop encodes the photon spectra emitted by Cherenkov radiation~\cite{Lin:2026nlr}. Together with the fact (made apparent by the preceding paragraph) that the evolution loop encodes collisional energy loss of a probe particle traversing a medium, its ability to describe radiative phenomena shows that a unified treatment of energy loss of fast-moving particles may be achieved without the need to separate each individual mechanism driving it --- which, depending on the strength of quantum interference effects, can be indispensable.

A field theorist might object that throughout this work we paid no attention to one of the most interesting features of field theories: their evolution under renormalization group flows. Indeed, except for specific symmetry-protected cases, any perturbative calculation of the evolution loop in a quantum field theory beyond leading order will encounter the usual regularization and renormalization considerations in its process of being carried out. The structure of the expansion we did in~\eqref{eq:Hamiltonian-expansion} and the power corrections that come with it are another aspect that we did not need to examine carefully here, but might prove equally interesting if new structural relations can be found from them.

It would be interesting to obtain connections with other recent developments in the theoretical study of dissipation and equilibration, such as 
new results on (generalized) fluctuation dissipation relations~\cite{Sieberer:2015hba,Tsuji:2016kep,Haehl:2017eob,Marchetto:2023xap,Zhang:2024gbx}, Schwinger-Keldysh effective field theory~\cite{Crossley:2015evo,Glorioso:2017fpd,Liu:2018kfw,Amoretti:2026bqg}, the framework of stochastic thermodynamics~\cite{Shiraishi2023,Falasco2025}, the thermalization process of QFTs more generally~\cite{Davison:2024msq,Delacretaz:2023ypv,Chen-Lin:2018kfl,Pappalardi:2021ahe,Delacretaz:2023pxm,Foini:2026pso,Qi:2026vht}, and specifically with recent developments in the understanding of ultracold atom experiments~\cite{Glidden:2020qmu,Huh:2023xso,Gazo:2023exc,Lannig:2023fzf,Martirosyan:2023mml} and far-from-equilibrium kinetic descriptions of heavy-ion collisions~\cite{Rajagopal:2024lou,Rajagopal:2025nca,Brewer:2019oha,Brewer:2022vkq,Mikheev:2022fdl,DeLescluze:2025gaa,Berges:2025ccd,DuPlessis:2026qjy,Heller:2023mah,Preis:2022uqs}.
Related developments in holography~\cite{Giataganas:2018ekx,Glorioso:2018mmw,Jana:2020vyx,Pantelidou:2022ftm}, together with the recent result for the complete momentum change probability of a heavy quark in terms of a string theory calculation~\cite{Rajagopal:2025ukd}, also open natural new directions of inquiry and suggest generalizations beyond the constant velocity approximation employed so far.

A concrete condensed matter system in which our results may have direct applicability is that of heavy impurities in quantum gases, specifically the Fermi polaron~\cite{Hennebichler:2026tqb,Rautenberg:2026lba,Grusdt_2025,Wang2023,Scazza2022,Schmidt_2018}, where a single heavy impurity interacts with excitations of a surrounding quantum environment. It would be interesting to find connections; more concretely, to find which observables would be interesting to calculate using the formalism we have introduced here.

Our results may also be of interest beyond setups where a probe particle limit is a natural approximation to the dynamics. The stochastic dynamics that emerges from treating the (observable) Universe as an open quantum system is a topic of active research in the context of inflation~\cite{DaddiHammou:2022itk,Salcedo:2024smn,Colas:2024lse,deKruijf:2024ufs,Lopez:2025arw,Li:2025azq,Li:2026lwl,Cespedes:2026fdp,Green:2026nnw}.
While written in terms of other variables, such open dynamics may nonetheless exhibit symmetries that are analog to the thermal KMS condition that underlies our result; it would be interesting to see whether concrete parallels could be drawn.

Given the breadth of fields for which the Langevin equation and its applications are relevant, we expect that this result will be of general interest.

\acknowledgments
I gratefully acknowledge inspiring conversations with Jean Du Plessis, Yue-Zhou Li, Joshua Lin, Ian Moult, Krishna Rajagopal, Soo-Jong Rey, Sarah E. Shandera, Urs A. Wiedemann, and Xiaojun Yao.
This research was supported in part by grant NSF PHY-2309135 to the Kavli Institute for Theoretical Physics (KITP) and by grant 994312 from the Simons Foundation. This research has benefited from the Mitchel Postdoctoral Scholar Career Development Fund at KITP. This work was initiated at the Institute for Nuclear Theory (INT) at the University of Washington, during the program ``Open Quantum Systems: Dissipative Dynamics from Quarks to the Cosmos'' INT-25-3b, which I thank for its kind hospitality and stimulating research environment. This research was supported in part by the INT's U.S. Department of Energy grant No. DE-FG02-00ER41132. This work was performed in part at the Aspen Center for Physics, which is supported by National Science Foundation grant PHY-2210452 and by a grant from the Simons Foundation (1161654, Troyer).

\appendix

\section{The late time limit of the evolution loop} \label{app:large-t-derivation}

In this Appendix we discuss how Eq.~\eqref{eq:markovian} can be derived. We start from the explicit expression
\begin{equation}
    \langle W^{\vg,t}_{il,jk} \rangle (\Lb) = \frac{{\rm Tr}_{\Hphi} \! \left\{ \left[ \bar{\rm P} \exp \left( i \int_{0}^{t} dt' \mathcal{O} \! \left( \vg t' + \Lb  ,t' \right) \right) \right]_{kl} \left[ {\rm P} \exp \left( -i \int_{0}^{t} dt' \mathcal{O} \! \left( \vg t' ,t' \right) \right) \right]_{ij} e^{-\beta H_{\varphi} }  \right\} }{ {\rm Tr}_{\Hphi} \! \left\{ e^{-\beta H_{\varphi} }  \right\}  } \, , \label{eq:evol-op-appendix}
\end{equation}
and proceed to convert the trace into a path integral by inserting a complete basis of states at each point in time.

To preserve a certain degree of familiarity with the notation used for textbook field theory path integrals, we will insert identity matrices written as complete bases of states in terms of the eigenstates of some family of fields $\varphi(\x)$ that fully describe the environment, and assume that we can write $\Op$ in terms of the $\varphi$ fields as a functional thereof at equal time
\begin{equation}
    \Op(\x) = \Op[\varphi](\x)\, .
\end{equation}
After inserting the $\varphi$ eigenstate bases, the evolution loop~\eqref{eq:evol-op-appendix} becomes
\begin{align}
    \langle W^{\vg,t}_{il,jk} \rangle (\Lb) &= \frac{1}{Z} \int D\varphi_1 D \varphi_2 D\varphi_E e^{i S[\varphi_1]-iS[\varphi_2]-S_E[\varphi_E]} \nonumber \\ & \quad \quad \,\, \left[ \bar{\rm P} \exp \left( i \int_{0}^{t} dt' \mathcal{O}[\varphi_2] \! \left( \vg t' + \Lb  ,t' \right) \right) \right]_{kl} \left[ {\rm P} \exp \left( -i \int_{0}^{t} dt' \mathcal{O}[\varphi_1] \! \left( \vg t' ,t' \right) \right) \right]_{ij}  \, , \label{eq:evol-loop-appendix-2}
\end{align}
which is the familiar Schwinger-Keldysh path integral representation of time evolution acting on a thermal density matrix, where we have omitted the matching conditions that the field values must satisfy at the respective end of their time integration domains, which can be found in thermal field theory textbooks~\cite{Bellac:2011kqa,Kapusta:2006pm,Laine:2016hma}. (We also stress that $S[\varphi]$ is not to be confused with $S(\Lb,\vg)$.) The normalization is simply given by $Z = \int D \varphi_E e^{-S_E[\varphi_E]}$. The path ordering symbols ${\rm P}$ and $\bar{\rm P}$ indicate that the operators $\Op$ (viewed as matrices acting on the internal labels $j,l$) are ordered from right to left from the lower integration limit to the upper integration limit if they are path-ordered (${\rm P}$) and from left to right if they are anti-path-ordered ($\bar{\rm P}$). This naturally matches the operator ordering in the quantum theory that determines this path integral.

It is instructive to work out the case where the internal labels can only take one value, i.e., the case where, for any given configuration $\varphi$, $\Op[\varphi]$ takes numerical values (in contrast to the general case where it is matrix-valued). In this case, the path-ordering symbols may be dropped, and the evolution loop~\eqref{eq:evol-loop-appendix-2} can be written as
\begin{align}
    \langle W^{\vg,t} \rangle (\Lb) &= \left\langle \exp \left( i \int_{0}^{t} dt' \mathcal{O}[\varphi_2] \! \left( \vg t' + \Lb  ,t' \right) \right) \exp \left( -i \int_{0}^{t} dt' \mathcal{O}[\varphi_1] \! \left( \vg t' ,t' \right) \right)\right\rangle_\varphi \nonumber \\
    &= \left\langle \exp \left( -i \int_{0}^{2t} dt' \mathcal{O}_{\Lb,\vg}[\varphi_c] ( t' ) \right)\right\rangle_\varphi \, , \label{eq:evol-loop-abelian-onepath}
\end{align}
where we have introduced
\begin{equation}
    \mathcal{O}_{\Lb,\vg}[\varphi_c] ( t' ) = \begin{cases} \mathcal{O}[\varphi_1] \! \left( \vg t' ,t' \right) & {\rm if} \,\, t' < t \, , \\  -\mathcal{O}[\varphi_2] \! \left( \vg (2t-t') + \Lb  ,2t-t' \right) & {\rm if} \,\, t' > t \, ,
    \end{cases}
\end{equation}
and the average $\langle \cdot \rangle_\varphi$ as
\begin{equation}
    \left\langle A \right\rangle_\varphi = \int D\varphi_1 D \varphi_2 D\varphi_E e^{i S[\varphi_1]-iS[\varphi_2]-S_E[\varphi_E]} A \, , \label{eq:phi-average}
\end{equation}
where we again emphasize that $S[\varphi]$ is not to be confused with $S(\Lb,\vg)$.

This is exactly amenable to employ a cumulant expansion~\cite{Kubo:1962dyl}. It follows that
\begin{align}
    \langle W^{\vg,t} \rangle (\Lb) &= \exp \left( \sum_{n=1}^\infty (-i)^n \int_0^{2t} dt_n \ldots \int_0^{t_2} dt_1 \langle \mathcal{O}_{\Lb,\vg}[\varphi_c] ( t_n ) \ldots \mathcal{O}_{\Lb,\vg}[\varphi_c] ( t_1 ) \rangle_{\varphi, c} \right) \, , \label{eq:cumulant-exp}
\end{align}
where the subscript $c$ stands for ``connected'' or ``cumulant.'' The reason to call them ``connected'' is that in any perturbative calculation around a (infinite-dimensional) Gaussian measure, as is the case in textbook quantum field theory (e.g., chapter 9 of~\cite{Srednicki:2007qs}), the Feynman diagrams that represent these contributions are themselves connected diagrams. Rearranging the time integrals and restoring our original notation for the operators $\Op$, we see that
\begin{align}
    \langle W^{\vg,t} \rangle (\Lb) &= \exp \Bigg( \sum_{\substack{n=0 , m = 0 \\ n+m>0}}^\infty \frac{i^m (-i)^n}{m!n!} \int_0^{t} dt_m' \ldots \int_0^{t} dt_1' \int_0^{t} dt_n \ldots \int_0^{t} dt_1 \nonumber \\  & \quad\quad\quad\quad \langle \bar{\mathcal{T}} \{\Op( \vg t_1' + \Lb, t_1' ) \cdots \Op(\vg t_m' + \Lb, t_m' )\} \mathcal{T}\{\Op(\vg t_n, t_n) \cdots \Op(\vg t_1, t_1) \} \rangle_{c} \Bigg) \, , \label{eq:evol-loop-abelian-exponentiated}
\end{align}
where $\mathcal{T}$, $\bar{\mathcal{T}}$ are time-ordering and anti-time-ordering symbols, respectively.

The existence of a Markovian limit~\eqref{eq:markovian} follows from the shift symmetry of the integrand in~\eqref{eq:evol-loop-abelian-exponentiated} that takes $(\{t_i\}_i,\{t_j'\}_j) \to(\{t_i + \Delta t\}_i,\{t_j' + \Delta t\}_j) $. \textit{Provided} there exists a scale in the theory such that correlations die out at large time separations, the time integrals may be evaluated in the $t \to \infty$ limit as
\begin{align}
    &\int_0^{t} dt_m' \ldots \int_0^{t} dt_1' \int_0^{t} dt_n \ldots \int_0^{t} dt_1 \nonumber \\  & \quad \langle \bar{\mathcal{T}} \{\Op( \vg t_1' + \Lb, t_1' ) \cdots \Op(\vg t_m' + \Lb, t_m' )\} \mathcal{T}\{\Op(\vg t_n, t_n) \cdots \Op(\vg t_1, t_1) \} \rangle_{c} \nonumber \\
    \overset{t \to \infty}{\longrightarrow} \quad & t \int_{-\infty}^\infty dt_m' \ldots \int_{-\infty}^\infty dt_1' \int_{-\infty}^\infty dt_n \ldots \int_{-\infty}^\infty dt_2 \nonumber \\  & \quad \langle \bar{\mathcal{T}} \{\Op( \vg t_1' + \Lb, t_1' ) \cdots \Op(\vg t_m' + \Lb, t_m' )\} \mathcal{T}\{\Op(\vg t_n, t_n) \cdots \Op({\bf 0}, 0) \} \rangle_{c} \, , \label{eq:late-time-limit}
\end{align}
thus implying that $\partial_t \langle W^{\vg,t} \rangle (\Lb) = - S(\Lb,\vg) \langle W^{\vg,t} \rangle (\Lb)$ as in~\eqref{eq:markovian}, with
\begin{align}
    S(\Lb,\vg) = - \sum_{\substack{n=0 , m = 0 \\ n+m>0}}^\infty & \frac{i^m (-i)^n}{m!n!} \int_{-\infty}^\infty  dt_m' \ldots dt_1'  dt_n \ldots dt_2 \nonumber \\ & \langle \bar{\mathcal{T}} \{\Op( \vg t_1' + \Lb, t_1' ) \cdots \Op(\vg t_m' + \Lb, t_m' )\} \mathcal{T}\{\Op(\vg t_n, t_n) \cdots \Op({\bf 0}, 0) \} \rangle_{c} \, .
\end{align}
We thus have our desired result.

The case where $\Op$ is matrix-valued may be examined in a similar manner by writing the path-ordered exponentials in terms of the eigenvalues and eigenvectors of the $\Op$ matrices for each field configuration, which leads to exponentiated cumulants of the eigenvalues of $\Op$ (which are functionals of $\varphi$), with a measure that contains the weights in~\eqref{eq:phi-average} and the inner product between the eigenstates of $\Op$ (which are typically not orthogonal, because they are evaluated on successive field configurations that are generally different from each other). The resulting measure is matrix-valued, and each of its entries can be evaluated using the cumulant expansion for the statistics of the eigenvalues of $\Op$ (this is most conveniently done if one already knows the basis in which the evolution loop is diagonal, so that scalar averages may be defined by projecting onto each eigenvector separately; otherwise one would have to find a matrix-valued normalization for the average that ensures that the expectation value of the identity operator is one, consistent with the discussion in~\cite{Kubo:1962dyl}). By the same token as before, if a scale is present in the theory that dampens out correlations between the eigenvalues of $\Op$ at large time separation, the time dependence in the exponent of the cumulant expansion will be linear, and we thus arrive at the desired conclusion.

In practice, however, the above considerations do not replace a calculation; they only provide structural guidance. In order to know at what time, or even whether, a Markovian regime emerges, one needs to establish that a scale in the theory describing the environment indeed generates correlations that decay sufficiently rapidly for the above discussion to be relevant.

\bibliography{main.bib}

\end{document}